\documentclass[aps,prx,11pt,superscriptaddress, twocolumn, tightenlines,nobibnotes,notitlepage]{revtex4-1}

\usepackage{amsmath}    
\usepackage{graphicx}   
\usepackage{verbatim}   
\usepackage{color}      
\usepackage{subfigure}  
\usepackage[margin=0.63in]{geometry}
\usepackage{soul}
\usepackage{bm}
\usepackage{gensymb}
\usepackage{nicefrac}
\usepackage{hyperref}   
\hypersetup{
	colorlinks=true,
	linkcolor={blue},
	citecolor={blue},
	urlcolor={blue}
}
\setcitestyle{super}
\usepackage[all]{hypcap}
\makeatletter
\renewcommand\frontmatter@abstractwidth{\dimexpr\textwidth\relax}
\makeatother

\makeatletter
\renewcommand*{\fnum@figure}{{\normalfont\bfseries Fig.~\thefigure}}
\renewcommand*{\@caption@fignum@sep}{\textbf{ \vrule width 1pt }}
\makeatother

\newcommand{\minus}{\scalebox{0.75}[1.0]{$-$}}

\begin{document} 
\setlength{\abovedisplayskip}{10pt}
\setlength{\belowdisplayskip}{10pt}

\title{\LARGE Tuneable Correlated Disorder in Alloys}
\author{D. Chaney}
\email{daniel.chaney@bristol.ac.uk}
\affiliation{School of Physics, University of Bristol, Tyndall Avenue, Bristol, BS8 1TL, UK}
\affiliation{European Synchrotron Radiation Facility, BP 220, F-38043 Grenoble, France}
\author{A. Castellano}
\affiliation{CEA, DAM, DIF, F-91297 Arpajon, France}
\author{A. Bosak}
\affiliation{European Synchrotron Radiation Facility, BP 220, F-38043 Grenoble, France}
\author{J. Bouchet}
\author{F. Bottin}
\author{B. Dorado}
\affiliation{CEA, DAM, DIF, F-91297 Arpajon, France}
\author{L. Paolasini}
\affiliation{European Synchrotron Radiation Facility, BP 220, F-38043 Grenoble, France}
\author{S. Rennie}
\author{C. Bell}
\author{R. Springell}
\email{phrss@bristol.ac.uk}
\author{G. H. Lander}
\affiliation{School of Physics, University of Bristol, Tyndall Avenue, Bristol, BS8 1TL, UK}

\date{\today}

\begin{abstract}
\noindent\textbf{Understanding the role of disorder and the correlations that exist within it, is one of the defining challenges in contemporary materials science. However, there are few material systems, devoid of other complex interactions, which can be used to systematically study the effects of crystallographic conflict on correlated disorder. Here, we report extensive diffuse x-ray scattering studies on the epitaxially stabilised alloy $\mbox{U}_{1-x}\mbox{Mo}_x$, showing that a new form of intrinsically tuneable correlated disorder arises from a mismatch between the preferred symmetry of a crystallographic basis and the lattice upon which it is arranged. Furthermore, combining grazing incidence inelastic x-ray scattering and state-of-the-art \textit{ab initio} molecular dynamics simulations we discover strong disorder-phonon coupling. This breaks global symmetry and dramatically suppresses phonon-lifetimes compared to alloying alone, providing an additional design strategy for phonon engineering. These findings have implications wherever crystallographic conflict can be accommodated and may be exploited in the development of future functional materials.}
\end{abstract}  
\maketitle
\thispagestyle{plain}
\section{Introduction}
The periodicity imbued in a crystallographic lattice has been at the heart of condensed-matter science for over a century\cite{Bragg}. Deviations from perfect periodicity and the degree of randomness introduced\cite{100years,Diffuse Scattering} is often ignored by simple models or absorbed into mean-field approaches. However, there is rapidly growing understanding\cite{Nanoscale_Structure,Goodwin1,Goodwin2} that many phenomena may only be understood when one properly embraces the role of disorder and the correlations within it. Such correlations may be described as ‘locally periodic’, coupling to periodic material properties such as collective atomic vibrations, electronic states, etc\cite{Aperiodic}. In many functional materials, from leading ferroelectric\cite{ferro1,ferro2,ferro3} and thermoelectric candidates\cite{Bozin,PbTe,AgSbTe2} to photovoltaic perovskites\cite{photovoltaic} and ionic conductors\cite{ionic1}, correlated deviation from perfect periodicity plays a pivotal role in governing functionality. This is the purview of disorder engineering; controlling the disorder within a system to create materials with new or improved functionality\cite{functional}.

Most elements occupy high-symmetry structures (\textit{fcc}, \textit{hcp} or \textit{bcc}) where equidistant neighbours provide an isotropic local environment. However, a number of elements, for example P, S, Bi and Ga, crystallise with anisotropic neighbour distances. The intermediate actinides, U, Np and Pu, are an extreme example of this tendency, where a combination of orbital anisotropy and narrow $5f$ bandwidth $(1-3\:\mbox{eV})$ drives unique, low-symmetry structures\cite{Mettout,Soderlind}. For example, the groundstate of uranium $(\alpha\mbox{-phase})$ consists of zig-zag chains that facilitate different bond lengths\cite{Historical Persepctive}. 

By organising an anisotropic element, or more generally any anisotropic basis, onto a high-symmetry lattice an obvious conflict is created between the local environment preferred by the basis and the global symmetry imposed by the lattice. The central question is how the interplay between these two opposing characters is manifested in the crystal structure and ultimately the material properties. A subsequent question is to what extent any effects are tuneable. \textit{A priori} one may expect the formation of correlated disorder, which thrives on symmetry mismatch\cite{Goodwin1}, and the shallow configurational landscape facilitated by a high-symmetry lattice\cite{Goodwin2}. 

\begin{figure}[t!]
	\centering
	\includegraphics[width=0.98\linewidth]{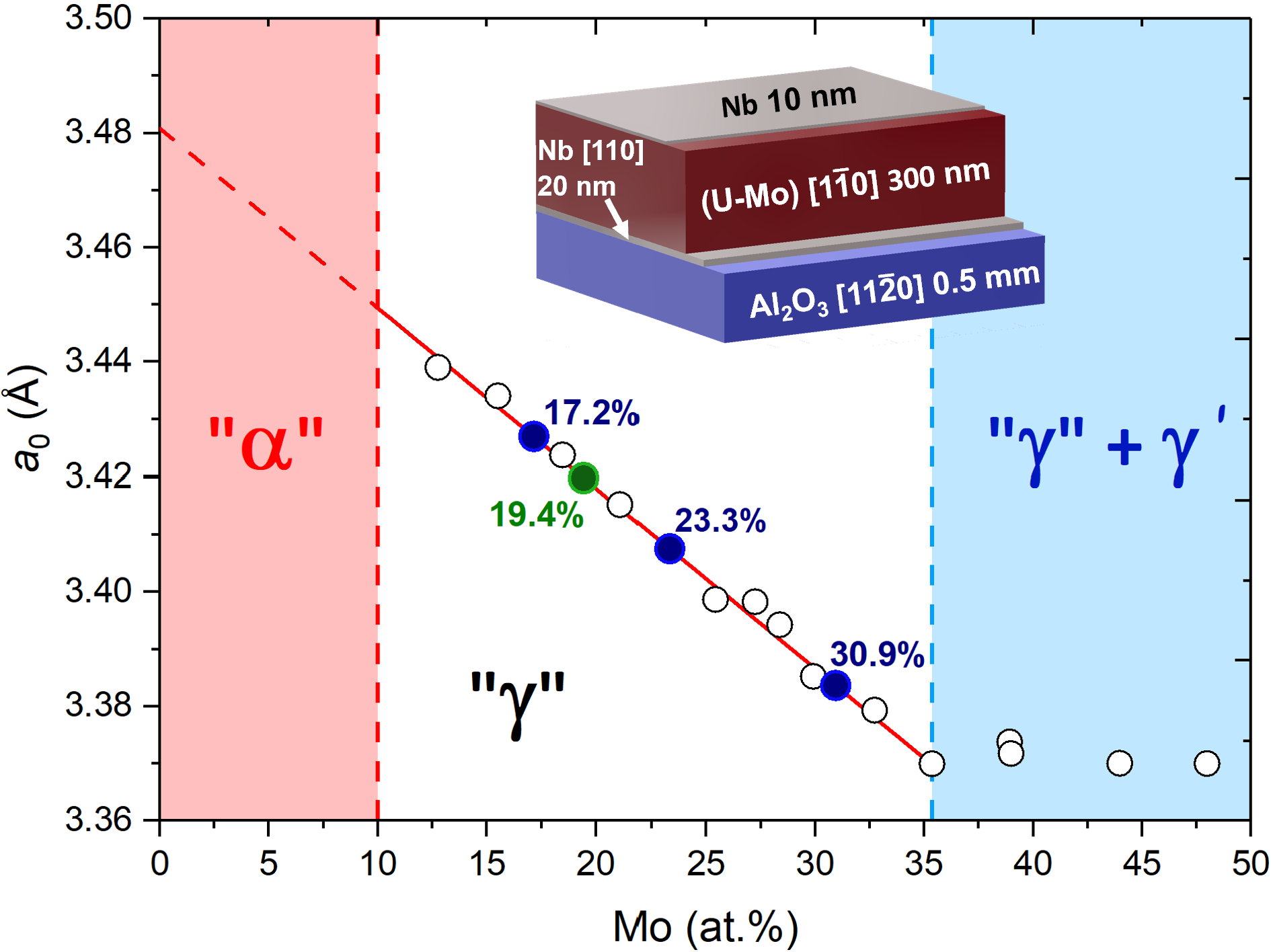}
	\caption{\textbf{ Nominal \textit{bcc} lattice parameter, $\bm{a_{\mbox{\tiny 0}}}$, as a function of at.\%. Mo} $a_{\mbox{\tiny 0}}$ was determined from the lattice parameter using the empirical fit (red line) to the data from Dwight\cite{Dwight} (open circles). $\alpha$, $\gamma$ and phase separation fields are highlighted in red, white and blue, respectively. The first two contain all metastable phases with the indicated parent structure. The latter contains metastable $\gamma$ along with the $\gamma^{'}$ phase, the tetragonal compound $\mbox{U}_2\mbox{Mo}$. $300\:\mbox{nm}$ and $70\:\mbox{nm}$ films are shown as blue and green solid points, respectively. (Insert) A schematic of the film structure.}
	\label{Fig1}
\end{figure}

When addressing these challenges, \textit{pseudo-bcc} $\left(\gamma\right)$  uranium – transition metal alloys are exemplars due to the considerable symmetry mismatch between the \textit{bcc} lattice and the orthorhombic \textit{Cmcm} groundstate of $\alpha\mbox{-uranium}$\cite{Historical Persepctive}. It is also already known that, within the $\gamma\mbox{-field}$ (Fig.\ref{Fig1}), these systems show evidence of locally distorted structures at ambient conditions\cite{Yakel}. This fact is often missed\cite{LLNL,Monolithic,Lopes} as the characteristic signal of local correlations are low intensity and only accessible via diffuse scattering methods\cite{Goodwin1,Diffuse Scattering,Total}. Crucially, an in-depth, systematic study of local correlations in such systems and their impact on material properties has been missing.

We have stabilised a series of $\mbox{U}_{1-x}\mbox{Mo}_{x}$ single-crystal thin films via epitaxial matching\cite{Adamska}, designed to facilitate a high degree of tunability. With extensive diffuse x-ray scattering studies, we show that the conflict created by a mismatch in preferred symmetry between a basis and the lattice produces a new form of intrinsically tuneable correlated disorder where every atom is displaced, lowering local symmetry, while maintaining the higher average symmetry imposed by the lattice. Through grazing incidence inelastic x-ray scattering experiments, together with state-of-the-art \textit{ab initio} modelling, we show how this form of disorder couples to the phonon dispersion, ultimately affecting physical properties.

\section{Methods}

\subsection{Sample Synthesis}
All films in this study were produced by DC magnetron sputtering in a dedicated actinide sputtering chamber at the University of Bristol, UK. The film structure utilised was a refinement on previous work\cite{Adamska}. Firstly, a $20\:\mbox{nm}$, epitaxial $\left[110\right]$ Nb buffer was deposited onto commercially procured $1\:\mbox{cm}^{2}$ a-plane sapphire substrates at $600\:\degree\mbox{C}$. This epitaxial match is well documented\cite{match} and produces high quality, single domain, $\left[110\right]$ Nb layers which act as both a chemical buffer, protecting oxygen diffusing from the sapphire substrate into the uranium layer, and as an epitaxial match to $\gamma\mbox{-U}$. A $300\;\mbox{nm}$ alloy layer was then formed by uranium-molybdenum co-deposition at $800\:\degree\mbox{C}$ where the alloy composition was tuned by adjusting the relative sputtering rates of each material. Films were probed with \textit{in situ} reflection high energy electron diffraction (RHEED) between stages to check for angle dependant signal characteristic of epitaxial, single crystal films. Finally, all films were capped at room temperature with 15 nm of polycrystalline Nb to prevent oxidation.

\subsection{In-house X-Ray Diffraction}	
All films were characterised via an in-house, copper source, Philips X’pert XRD. A total of $24$ specular and off-specular reflections were collected for each film and the lattice parameters were determined via linear least squared regression allowing the system orthorhombic freedom. Small tetragonality was observed in all samples such that the in-plane parameter was enlarged compared to the out-of-plane parameters evidencing small amounts of epitaxial clamping. This effect varies between a $0.8\%$ and $0.4\%$ increase of the in-plane parameter and is inversely proportional to Mo content. Accurate Mo content values were determined from the empirical curve shown in Figure \ref{Fig1} using a relaxed lattice parameter of the form: 

\begin{equation}
\frac{\sqrt{\frac{\left|\bm{a}\right|^2+\left|\bm{b}\right|^2}{2}}+\left|\bm{c}\right|}{2}
\end{equation}

\noindent where $\bm{c}$ lies in-plane and $\bm{a}$ and $\bm{b}$ are out-of-plane parameters. This is derived from the cubic requirement for $\left|\left[001\right]\right|=\left|\left[110\right]/\sqrt{2}\right|$ . Rocking curves confirmed a mosaic width of $0.5-1\degree$ for all samples. 

\subsection{Diffuse and Inelastic X-Ray Scattering}	
Both x-ray diffuse scattering and grazing incidence inelastic scattering (GI-IXS) measurements were performed at the $\mbox{ID}28$ beamline at the European Synchrotron Radiation Facility, Grenoble, France\cite{IXS,Diffuse} with all studies conducted at room temperature and pressure. The diffuse scattering studies utilised an incident beam energy of $12.65\:\mbox{keV}$ to move away from the uranium $L_3$ absorption edge. Data were collected at two detector $\left(2\theta\right)$ positions, $19\degree$ and $48\degree$, to access a sufficiently large portion of reciprocal space. Analysis was performed using a combination of in-house software and the CrysalisPro analysis suite\cite{Crysalis}. 

The GI-IXS technique is one we have used previously with epitaxial films of $\mbox{UO}_{2}$\cite{Rennie}. The data were collected over two experiments using $\mbox{Si}\left(999\right)$ and $\mbox{Si}\left(888\right)$ monochromator reflections, respectively. These settings correspond to incident beam energies of $17.794$ and $15.817\:\mbox{keV}$ with experimental resolutions of $3$ and $5.5\:\mbox{meV}$, respectively. The data were fitted with a Lorentzian component to describe the elastic signal and a damped harmonic oscillator fit for the inelastic components; all fitting functions were convoluted with the relevant resolution function. For the linewidth deconvolution procedure all non-instrumental broadening was assumed to have a Gaussian profile. The instrument has $9$ independent analysers that allows phonon energies to be determined over a small $\bm{q}$ range for one detector position. This accounts for the large point density on the phonon-dispersion curves, especially when the group velocity $\left(\frac{d\bm{q}}{dE}\right)$ is small. 

\subsection{Computational Modelling}	
The theoretical dispersion curves were calculated using density-functional theory with the ABINIT package\cite{ABINIT}. A full account of the theoretical methodology has been published by Castellano \textit{et al.}\cite{Castellano} and more discussion is given in supplemental material V\cite{SM}. 

\section{Probing the Local Order}

\begin{figure*}[t]
	\centering
	\includegraphics*[width=0.98\linewidth]{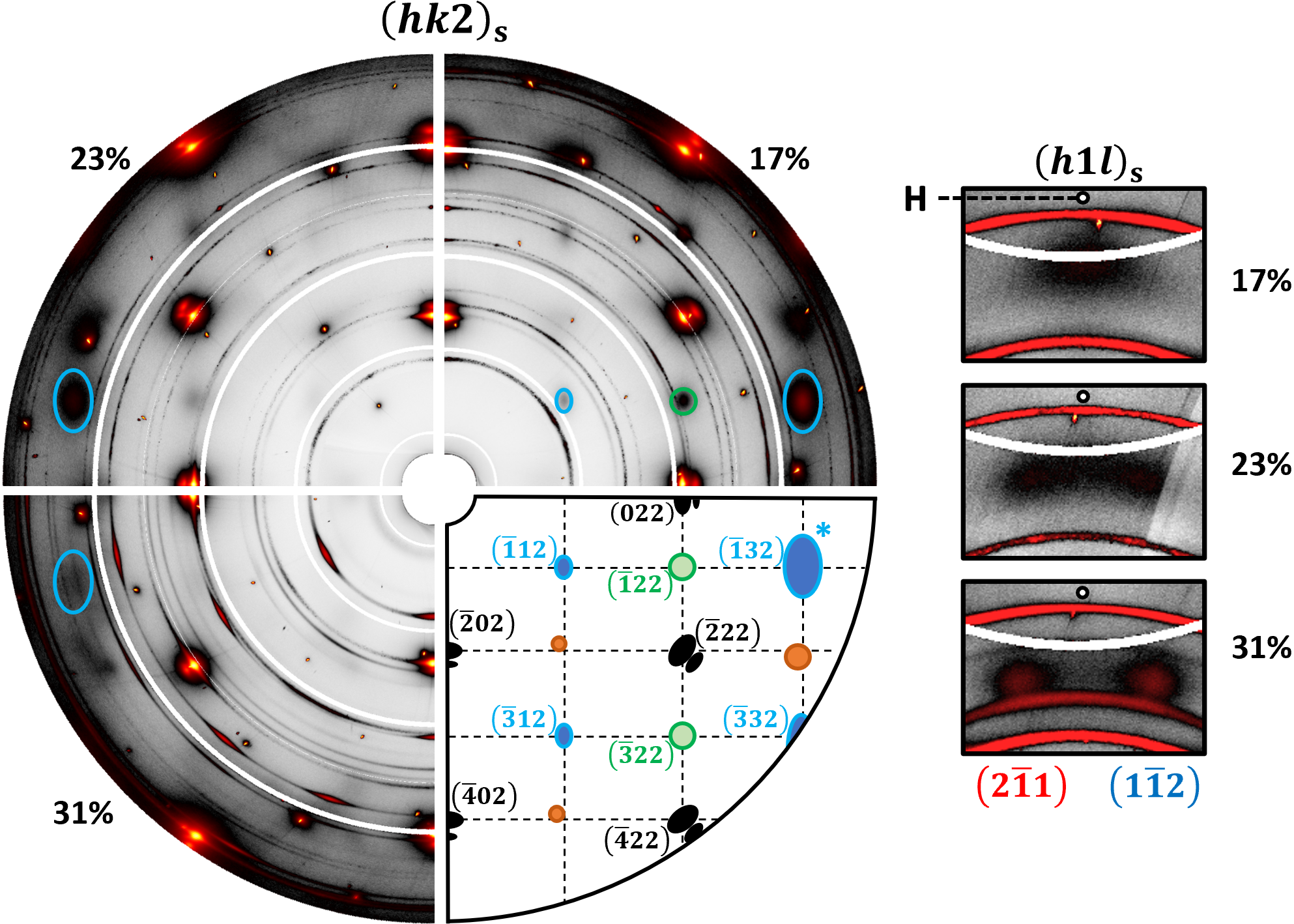}
	\caption{\textbf{ Reciprocal space reconstructions of the $\bm{(hk2)_{\mbox{\textbf{\scriptsize s}}}}$ (left) and $\bm{(h1l)}_{\mbox{\textbf{\scriptsize s}}}$ (right) planes} All three samples investigated are shown with relevant Mo at.\% indicated for each quadrant. A superstructure unit cell with $\bm{b}_{\mbox{\scriptsize s}}\equiv \bm{b}_{\mbox{\scriptsize p}}$ was used for all reconstructions. Reflections are categorized and indexed in the bottom right quadrant, with indices in superstructure notation. Reflections are indicated in black (parent Bragg peaks), blue (\textbf{N} – domain 1), green (\textbf{N} – domain 2) and orange (\textbf{H}). Nb buffer reflections appear as doublets outside of the parent \textit{bcc} reflections. Powder rings arise from the polycrystalline Nb cap and narrow intense peaks correspond to substrate Bragg reflections, neither of which are included schematically. The $(\bar{1}32)_{\mbox{\scriptsize s}}$ reflection, marked with an asterisk, is highlighted in all data sets to show mirror symmetry relations.
	(Right) Reconstructions for the $(h1l)_{\mbox{\scriptsize s}}$ plane highlighting the evolution of the \textbf{H} type signal. The \textbf{H} point is marked with a circle. $31 \:\mbox{at}.\%$ peaks are indexed with $\mbox{U}_2\mbox{Mo}$ with $\bm{c}$ parallel to $\bm{a}_{\mbox{\scriptsize p}}$ (red) and $\bm{c}_{\mbox{\scriptsize p}}$ (blue).} 
	\label{Fig2}
\end{figure*}

\subsection{Results from Diffuse X-Ray Scattering Studies}

A composite reciprocal space reconstruction, made from the three Mo concentrations investigated, is shown in Figure \ref{Fig2}. This clearly shows two types of diffuse reflections. The stronger of the two at the \textbf{N} positions, a translation of $\langle\nicefrac{1}{2},\nicefrac{1}{2},0\rangle_{\mbox{\scriptsize p}}$ from the \textit{bcc} Bragg positions, and the weaker near the \textbf{H} position, a translation of $\langle1,0,0\rangle_{\mbox{\scriptsize p}}$. Both diffuse scattering components were previously observed in a similar system\cite{Yakel} and were attributed to the same origin. However, with the aid of greatly increased flux, finer $\bm{q}\mbox{-resolution}$ and most importantly a systematic series of alloy concentrations, we show the diffuse signal comprises of two distinct sets, originating from different effects. The two types of diffuse signal show opposite trends with respect to Mo content, and will be referred to as \textbf{N} and \textbf{H}, respectively, based on their proximity to the corresponding symmetry positions in the \textit{bcc} Brillouin zone (BZ). Throughout this paper $\bm{Q}$ will refer to total momentum transfer such that $\bm{Q}=\bm{G}+\bm{q}$ where $\bm{G}$ is a reciprocal lattice vector and $\bm{q}$ is a vector within the relevant first BZ. Reciprocal space coordinates $(h,k,l)$ will be used throughout, with subscripts p and s indicating parent and superstructure properties, respectively.

\begin{figure}[t!]
	\centering
	\includegraphics*[width=0.98\linewidth]{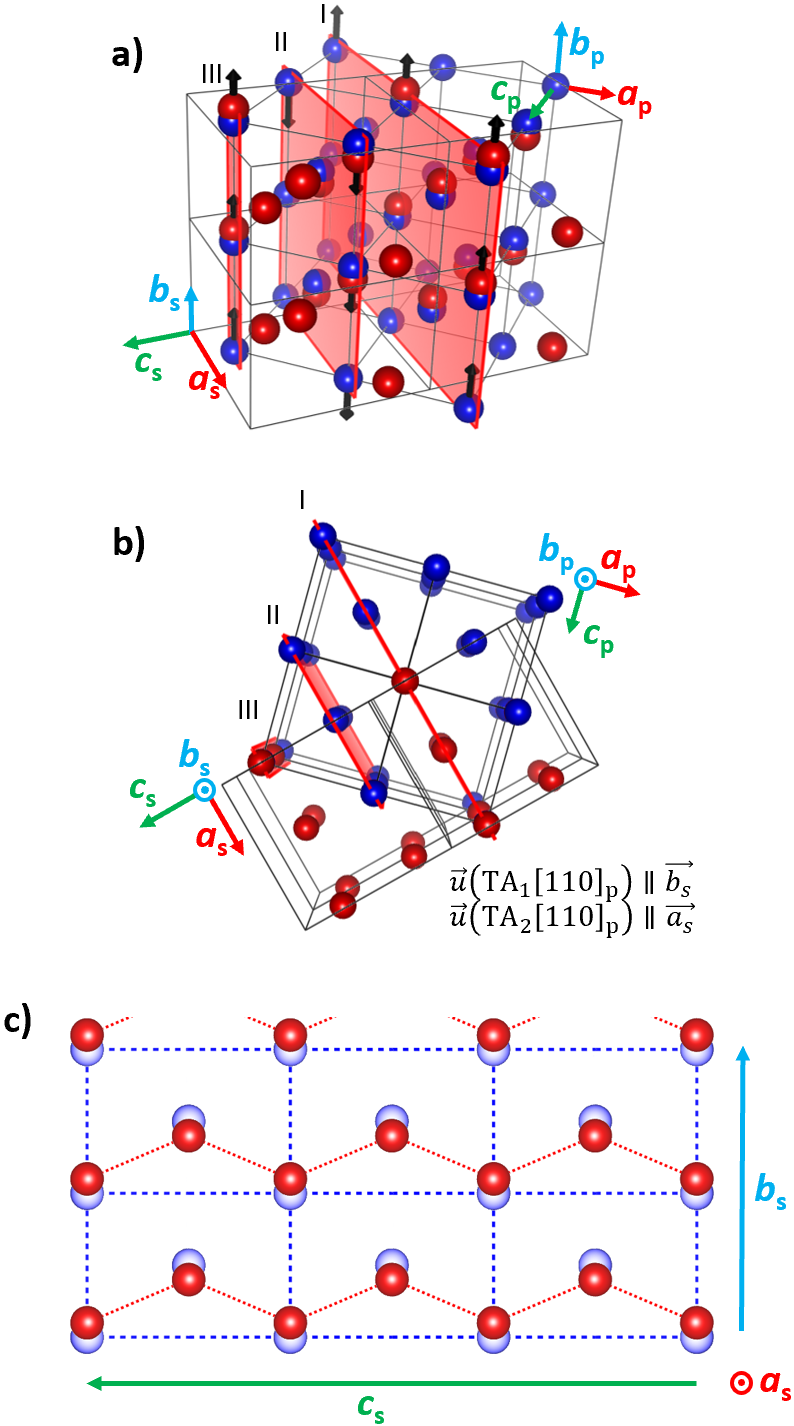}
	\caption{\textbf{ Relationship between parent and superstructures.} (a) Orientational relationship (see supplemental material IV\cite{SM} for transformation matrices) between parent (blue) and superstructure (red) unit cells for one of six possible domains, with $|\delta|=0.1$ for clarity, see main text for details. All atoms in the `shear plane' (highlighted red) move colinearly with direction of motion indicated by arrows on plane edge. Alternate planes, demarcated by I,I,III$\dots$, move in antiphase. (b) Top-down view showing the $45\degree$ relationship between the parent and superstructure. Directional relationship of $\mbox{TA}_{1}[110]_{\mbox{\scriptsize p}}$ and $\mbox{TA}_{2}[110]_{\mbox{\scriptsize p}}$ mode included bottom right. (c) Schematic of the atomic motions in a “phonon plane”. Blue dashed, and red dotted lines refer to interatomic bonding in the parent and superstructure unit cells, respectively.} 
	\label{Fig3}
\end{figure}

Considering the \textbf{H} reflections first, they occupy a lattice, close to but displaced from the \textbf{H} positions. There is also significant evolution with Mo concentration, as shown in the insert of Figure \ref{Fig2}, starting with one broad peak at low concentrations before splitting into multiple distinct peaks at the highest concentration. The final state is fully described by a coherent precipitate of $\mbox{U}_{2}\mbox{Mo}$ (see Fig. \ref{Fig1}), a chemically ordered, tetragonal compound $\left(c/a=2.87\right)$ known to precipitate above $33$ at.\% Mo\cite{U2Mo}. The signal in the two lower content samples is attributed to a precursor structure of $\mbox{U}_{2}\mbox{Mo}$, the exact nature of which is beyond the scope of this study.

Identification of the \textbf{H} point signal as precipitate, and not $\gamma$ phase reflections, as previously suggested\cite{Yakel}, allows us to isolate the second set of diffuse reflections, \textbf{N}, highlighted blue and green in Figure \ref{Fig2}. They are characterized by four key features: exact occupation of \textbf{N} positions, a prolate ellipsoidal shape, systematic absences at $\left(h0l\right)_{\mbox{\scriptsize s}}$ and a positive relationship between $k_{\mbox{\scriptsize s}}$ and diffuse intensity. All observed behaviours may be explained by a local structure formed by atomic displacements along $\langle010\rangle_{\mbox{\scriptsize s}}$ with no requirement for chemical ordering. The resulting structure recovers a \textit{Cmcm} symmetry, with anisotropic neighbour distances reminiscent of the uranium groundstate. The superstructure is defined by $\left|a_{\mbox{\scriptsize s}}\right|=\left|c_{\mbox{\scriptsize s}}\right|=\left|a_{\mbox{\scriptsize p}}\right|\sqrt{2}$ and $\left|b_{\mbox{\scriptsize s}}\right| = \left|a_{\mbox{\scriptsize p}}\right|$ with four atoms in the unit cell at $\pm\left(0,\nicefrac{1}{4}+\delta,\nicefrac{1}{4}\right)$ and $C\mbox{-face}$ centring assuring that $\left(h_{\mbox{\scriptsize s}} + k_{\mbox{\scriptsize s}}\right)$ must be even. The \textit{bcc} structure is recovered by setting the displacement magnitude $|\delta|=0$. Accounting for all possible domains, this structure describes all observed \textbf{N} reflections and constitutes a unique solution. We note that Ti exhibits a similar \textit{Cmcm} $\left(\delta\right)$ phase with a distorted \textit{bcc} structure at high pressure $\left(\mbox{above}\:140\:\mbox{GPa}\right)$\cite{delta-Ti}. However, the deviation from \textit{bcc} in $\delta\mbox{-Ti}$ is significantly more robust than the short-range correlations reported here. They find $\left|\delta\right|= 0.1$, compared to $0.01 - 0.03$ in our system, and $a/b$ and $c/b$ ratios significantly modified from $\sqrt{2}$.

The distortion which transforms from the parent to superstructure, shown in Figure \ref{Fig3}, may be described within a frozen phonon framework. The responsible mode being the $\mbox{TA}_{1}\left[110\right]_{\mbox{\scriptsize p}}$, with polarization along the fourfold axis $\left[001\right]_{\mbox{\scriptsize p}}$. Twelve-fold degeneracy provides six equivalent domains, which are all equally occupied as evidenced by equal intensity distribution, to within experimental uncertainty, between equivalent diffuse spots. Crystallographically, this degeneracy corresponds to the freedom to set $\bm{b}_{\mbox{\scriptsize s}}$ equivalent to any parent axis and for each choice there exists two possible orientational rotations, $\pm45\degree$ about $\bm{b}_{\mbox{\scriptsize s}}$. Accounting for time reversal this gives $12$ diffuse reflections about each parent Bragg position, one at every \textbf{N} position, for domains not shown in Figure \ref{Fig2} see supplemental material I\cite{SM}. 

Once mapped onto the superstructure lattice, the systematic absences and relationship between the diffuse intensity and momentum transfer may be easily understood by considering the structure factor at the \textbf{N} position as
\vspace*{-0.1mm}
\begin{equation}
\resizebox{0.43\textwidth}{!}{$\left|F_{\mbox{\textbf{\scriptsize N}}}\left(\bm{Q}\right)\right|\propto f(Q)\left|\sin\left[2\pi\left(\bm{Q\cdot u}\right)\right]\right| = f(Q)\left|\sin\left[2\pi\left(k_{\mbox{\scriptsize
			 s}}\delta\right)\right]\right|$},
\end{equation}

\noindent where $\bm{u}$ is the displacement vector and $f(Q)$ is the average atomic scattering factor. As $\bm{u}=\left[0\delta 0\right]_{\mbox{\scriptsize s}}$,  $\left|F_{\mbox{\textbf{\scriptsize N}}} \left(\bm{Q}\right)\right|$ couples directly to $k_{\mbox{\scriptsize s}}$ providing a positive intensity relationship with $k_{\mbox{\scriptsize s}}$ as well as systematic absences for $k_{\mbox{\scriptsize s}}=0$, as observed. For each domain pair $k_{\mbox{\scriptsize s}}$ is aligned with a different parent axis such that the true behaviour is hidden when treated within the parent representation. Since $\left|\delta \right|$ is small, the intensity reduction at parent Bragg positions is negligible, see supplemental material II\cite{SM}. It should be noted we observe no evidence for chemical ordering, which would display the reverse trend --- diffuse intensity decreasing with momentum transfer.

The local superstructure also provides a natural explanation for a prolate ellipsoidal diffuse signal. If one considers the physical meaning imbued in the superstructure axes; $\bm{b}_{\mbox{\scriptsize s}}$ and $\bm{c}_{\mbox{\scriptsize s}}$ are the directions of atomic motion and phonon propagation, respectively, which together compose a `phonon-plane', depicted in Figure \ref{Fig3}\textcolor{blue}{c}. The third axis, $\bm{a}_{\mbox{\scriptsize s}}$, is perpendicular to this plane and combined with $\bm{b}_{\mbox{\scriptsize s}}$, may be thought of as a `shear-plane', highlighted red in Figure \ref{Fig3}\textcolor{blue}{a}, within which all atomic motions are in-phase and adjacent planes perform a shearing motion. The long axis of the diffuse ellipsoid is aligned with $\bm{a}_{\mbox{\scriptsize s}}$ showing the correlations are significantly stronger within, as opposed to out of, the phonon plane. In the $17$ at.\% Mo system,  a lower limit on the correlation lengths $\left(\xi\right)$ may be extracted as $30\:\mbox{\AA}$ and $22\:\mbox{\AA}$ along $\bm{b}_{\mbox{\scriptsize s}}/\bm{c}_{\mbox{\scriptsize s}}$ and $\bm{a}_{\mbox{\scriptsize s}}$, respectively. Given the finite correlation lengths are the dominant broadening factor for the diffuse peaks these values are also expected to be a good approximation for the true values. Both the correlation lengths and the magnitude of individual displacements, proportional to inverse peak width and integrated area, respectively, display strong tunability with alloy composition. This is clearly demonstrated by the decrease in area and inverse width for samples with sequentially greater Mo content, evident in Figure \ref{Fig2} and explored further in supplemental material III\cite{SM}. Regarding the character of the correlations, given a large enough sample of $\bm{q}\mbox{-space}$, and a sufficiently good signal-noise ratio, it is possible to distinguish between embryonic wave-like distortions (Lorentzian) or nanodomains with constant $\left|\delta\right|$ (squared Lorentzian)\cite{Bosak}, however we do not observe a statistically different fit between the two functions for our data and as such the exact character remains undetermined.

By studying correlations in the alloy, we are also afforded an insight into the fundamental instability in pure \textit{bcc}-uranium $\left(\gamma\right)$, which is only stable above $1045\:\mbox{K}$\cite{Johann}. Therefore, we can confirm that the relevant room temperature instability is in the $\mbox{TA}_{1}\left[110\right]_{\mbox{\scriptsize p}}$ branch, evaluated at \textbf{N}, as predicted\cite{Soderlind2,Johann}. On cooling the uranium phase pathway passes from $\gamma$ through the highly complex $\beta\mbox{-phase}$\cite{Beta}, before transforming to the orthorhombic $\alpha\mbox{-phase}$. The exact $\gamma-\beta$ transformation mechanism is unclear. However, our study indicates that the $\mbox{TA}_{1}$ mode may play a significant, if not primary role.  This contrasts with the transition at higher pressures, where $\gamma$ transforms directly to $\alpha$ via the $\mbox{TA}_{2}\left[110\right]_{\mbox{\scriptsize p}}$ mode, atomic motion $\left[1\bar{1}0\right]_{\mbox{\scriptsize p}}$, as first proposed by Axe \textit{et al.}\cite{Axe}

\subsection{Discussion on Local Order}

Overall, the system should be thought of in terms of displacive disorder that lowers the local symmetry, with correlations governed by rules laid out within a frozen phonon model. The resulting local structure is twelve-fold degenerate, maintaining the higher average symmetry, while recovering anisotropic neighbour distances reminiscent of $\alpha$ uranium. Hence, it is clear that the intrinsic conflict created by a mismatch in preferred symmetry between a crystallographic basis and the lattice is resolved by the formation of correlated disorder. Globally, the high symmetry \textit{bcc} structure is preserved, whereas locally, a significant symmetry reduction is allowed. Importantly the observed state is not considered to be a disordered precursor to an ordered, low temperature state, as the lattice and basis symmetries are mutually incompatible. We believe this constitutes a new form of correlated disorder where every atom is displaced to form a short-range superstructure with no evidence of chemical order. 

Furthermore, this study highlights the important role that modern, high resolution, diffuse scattering has to play in contemporary materials physics. Recent instrumental advances have allowed us to build on earlier pioneering work\cite{Yakel} to fully understand the $\gamma\mbox{-UMo}$ system and we stress that, as other authors have recently highlighted\cite{Goodwin1,Goodwin2}, correlated disorder may be significantly more prevalent than previously thought. As such, diffuse scattering methods should be employed to investigate systems wherever correlations are suspected.

\begin{figure}[t]
	\centering
	\includegraphics*[width=\linewidth]{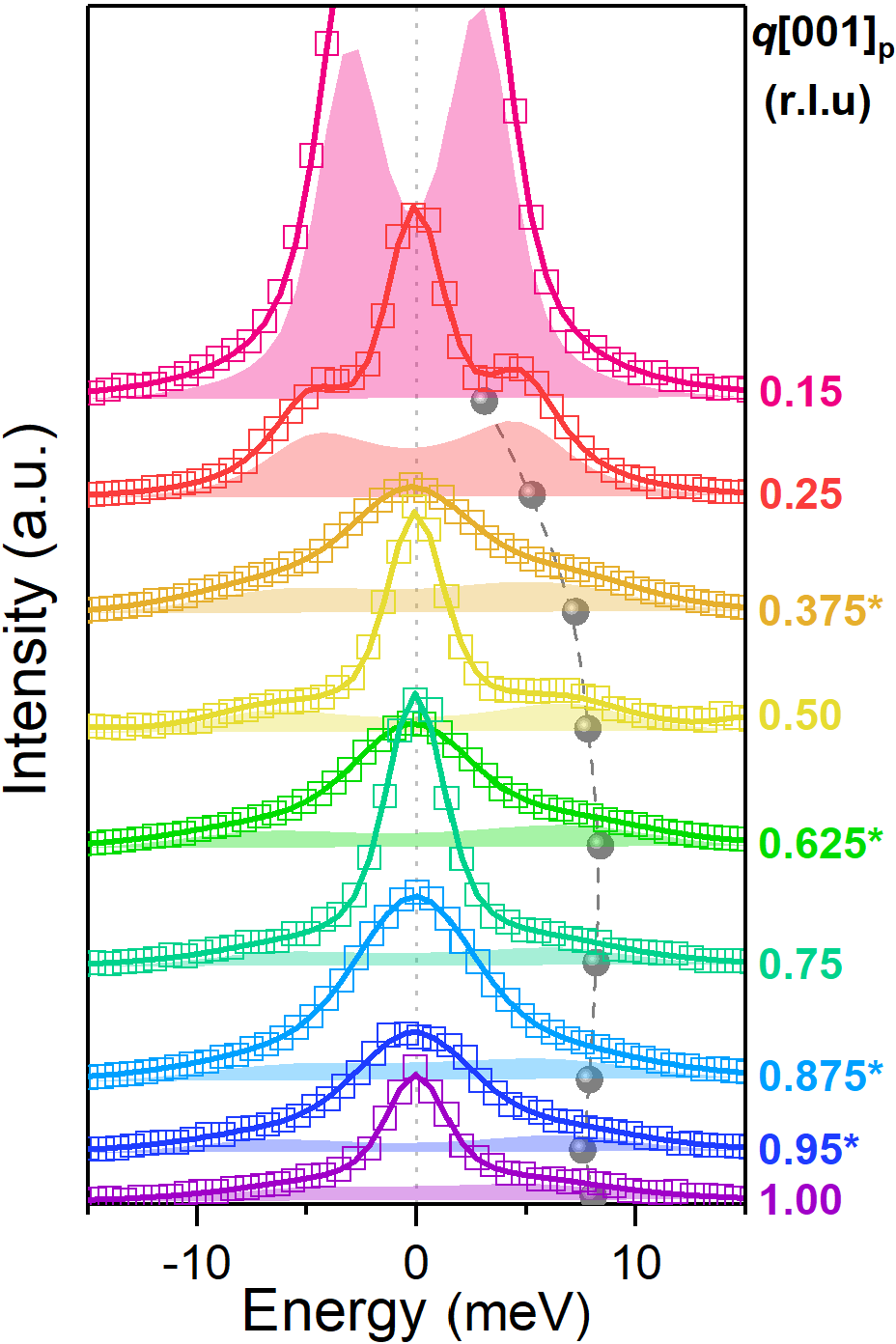}
	\caption{\textbf{ $\mbox{TA}\bm{[001]}_{\mbox{\scriptsize p}}$ intensity as a function of $\bm{q}$ along $\mathbf{\Gamma}\rightarrow\mathbf{H}$, for $\bm{23}\;\mbox{at}.\%$ Mo.} The strong elastic response at $\bm{q}=0.15$ has been omitted and the lower resolution (see methods) data, indicated by asterisks, are multiplied by 4 for clarity. Data points are shown as open squares, total fit as solid curves with the phonon contribution highlighted by shaded regions. Fitted phonon frequencies are projected on the plane as grey points with a grey dashed spline as a visual guide of the dispersion.} 
	\label{Fig4}
\end{figure}

\section{Assessing Disorder-Phonon Coupling}

\begin{figure*}[t!]
	\centering
	\includegraphics*[width=\linewidth]{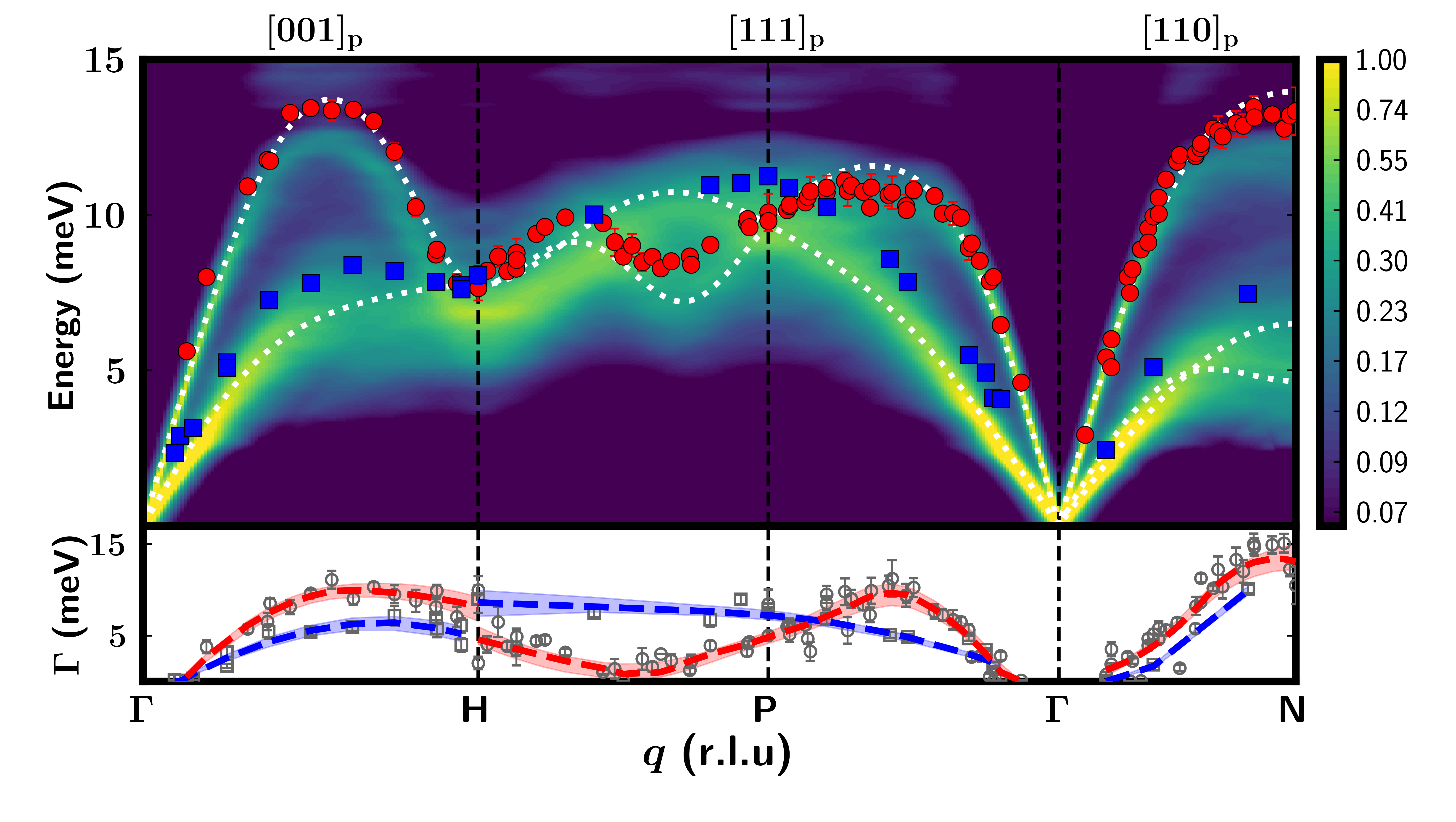}
	\caption{\textbf{ Phonon energy and linewidth dispersions.} (Top Panel) Experimental (23 at.\% Mo) and theoretical (25 at.\% Mo) phonon dispersion curves. Transverse (longitudinal) acoustic modes are shown as blue squares (red circles) and theoretical results from a virtual crystal approximation are shown as dashed white lines. The full spectral function is plotted as a $\log_{0.6}$ colour map to rescale the intensity divergence at gamma. All directions are within the parent BZ. (Bottom Panel) Raw linewidths, $\Gamma_{\mbox{\footnotesize o}}$, are shown as grey squares (TA) and circles (LA) with deconvoluted linewidths $\Gamma_{\mbox{\footnotesize d}}$ shown by  dashed blue (TA) and red (LA) trendlines. The smoothing methodology is described in in the main body of the text. Errors were determined by smoothing raw errors through the same algorithm and are shown as confidence bands.}
	\label{Fig5}
\end{figure*}

\subsection{Results from Inelastic X-Ray Scattering Studies and \textit{ab initio} Modelling}

As the underlying periodicity of a system is modified by the presence of correlated disorder one may reasonably expect to observe fingerprints of correlations among phenomena that are also periodic in nature. As such we have studied phonon dispersions for one sample $\left(23\:\mbox{at.\% Mo}\right)$, to investigate the effect of correlated static distortions on the dynamic properties. The data were collected in grazing-incidence inelastic x-ray scattering studies\cite{GRIX,Rennie}. Representative constant $\bm{Q}$ scans along $\bm{q}=\left[001\right]_{\mbox{\scriptsize p}}$ are shown in Figure \ref{Fig4} and the full dispersion in Figure \ref{Fig5}.

Our most important result is that, apart from close to the zone centres $\left(\Gamma\right)$, we observe considerable broadening in the phonon linewidths, which are inversely proportional to lattice thermal conductivity, as can be seen in Figure \ref{Fig4}. This is consistent with the observations of Brubaker \textit{et al.} who report maximum linewidths of $\sim20\:\mbox{meV}$ at \textbf{N} for the $20\;\mbox{at.\% Mo}$ system\cite{LLNL} compared with a $\sim14\:\mbox{meV}$ maximum observed in this study. The comparative decrease is unsurprising given their study was conducted at a lower alloying percentage and the strength of the correlated disorder is expected to increase correspondingly. Similar effects have been observed in $\omega$ phase alloys\cite{omega,omega2}, however in such systems broadening is observed for only part of the dispersion and is attributed to embryonic  regions that provide extra observable branches. The broadening in our system may be attributed to two main factors; alloy related effects and the presence of short-range correlations discussed above. 

In order to isolate the effects of short-range correlations and therefore assess their relative importance, modelling was performed via \textit{ab initio} molecular dynamic simulations, a detailed discussion of which is given in supplemental material V, VI\cite{SM} and by Castellano \textit{et al.}\cite{Castellano} The resulting spectral function, shown as a colour map in Figure \ref{Fig5}, includes broadening from mass difference as well as interatomic force constants' (IFC) dependent on both bond type and length, the latter providing an approximation for random displacive disorder. Anharmonicity from phonon-phonon interactions (at $300\:\mbox{K}$) was assessed to be much less than $1\:\mbox{meV}$ and phonon-point defect scattering was assumed to be negligible. As such, the simulation is assumed to capture all major alloy related broadening, but no correlation effects. This allows us to approximate the deconvoluted linewidth shown in the lower panel of Figure \ref{Fig5} by subtracting in quadrature the modelled width, $\Gamma_{\mbox{\footnotesize m}}$, from observed widths, $\Gamma_{\mbox{\footnotesize o}}$. The deconvoluted linewidths, $\Gamma_{\mbox{\footnotesize d}}$, show a strong dependence on phonon branch, direction and energy, discussed further below. However, as evidenced by the excellent agreement between the deconvoluted curves and experimental data, the alloy contribution is comparatively small for almost the entirety of the dispersion. The theoretically predicted linewidths, which include only alloying effects, never exceed $2\:\mbox{meV}$. Assuming the validity of the quadratic approximation, $\Gamma_{\mbox{\footnotesize d}} = \sqrt{\Gamma_{\mbox{\footnotesize o}}^2\: \minus \: \Gamma_{\mbox{\footnotesize m}}^2}$, we conclude that, barring the exceptional positions discussed below, the majority of the observed broadening may be attributed to the presence of correlated displacive disorder.

\begin{figure}[t]
	\centering
	\includegraphics*[width=\linewidth]{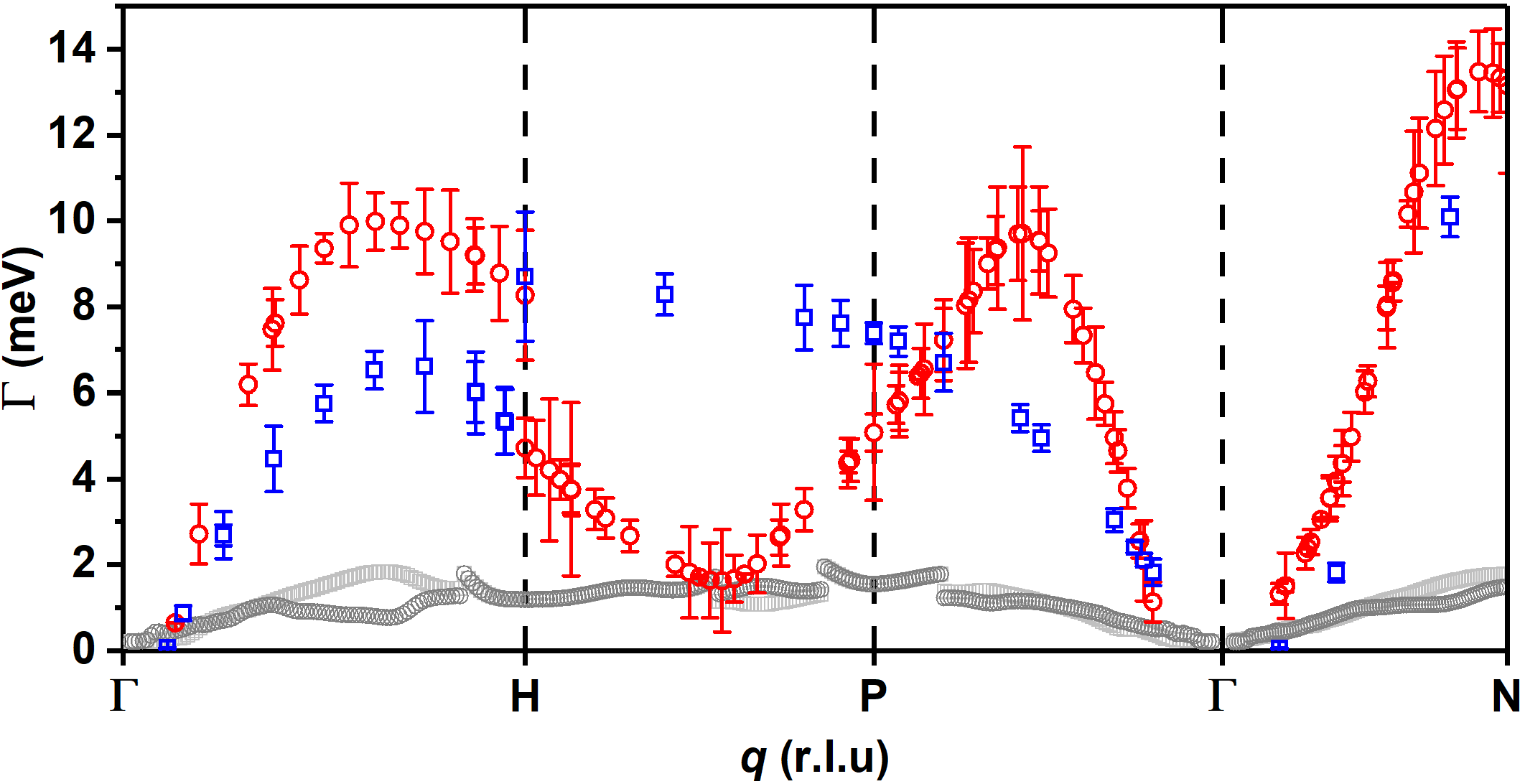}
	\caption{\textbf{ Comparison of theoretical and experimental linewidth dispersions.} Smoothed experimental LA and TA linewidths, $\Gamma_{\mbox{\footnotesize o}}$, shown as open red circles and blue squares, respectively. Error bars are taken directly from the unsmoothed data and were determined from an in-house fitting algorithm. Theoretical LA and TA linewidths, $\Gamma_{\mbox{\footnotesize m}}$, shown as open dark grey circles and light grey squares, respectively.} 
	\label{FigS4}
\end{figure}

To further probe the $\bm{q}\mbox{-dependence}$ in the disorder-phonon coupling, we produce a linewidth dispersion, Figure \ref{FigS4}, directly comparing $\Gamma_{\mbox{\footnotesize o}}$ and $\Gamma_{\mbox{\footnotesize m}}$. The experimental data have been smoothed using a locally estimated scatterplot smoothing algorithm to account for the variance introduced when fitting very broad, low intensity phonons and extract a trend from the raw linewidths shown in the lower panel of Fig. \ref{Fig5}. The smoothing algorithm was applied separately for each branch and each crystallographic direction. An identical smoothing methodology was used to produce the $\Gamma_{\mbox{\footnotesize d}}$ trendlines and errors shown in the lower panel of Figure \ref{Fig5}.  

Firstly, we note that, as is also clear from Figure \ref{Fig5}, the branch identity is not preserved through the change of direction that occurs at \textbf{H}. LA linewidths become TA linewidths and vice versa. Such a discontinuity is allowed since phonon lifetimes are a third-order term and have no requirement to vary smoothly through high symmetry positions. Secondly, as stated above, it can be clearly seen that for almost the entirety of the dispersion the measured linewidths are greatly in-excess of those predicted from alloying alone, allowing us to attribute the observed broadening to the presence of correlated disorder. However, there are two places were this is not true, firstly, near the BZ centre and, secondly, close to the $\mbox{LA-}2/3\langle111\rangle_{\mbox{\scriptsize p}}$ position. The first case is a consequence of wavelength; collective vibrations are insensitive to structural modulations with characteristic length scales shorter than their wavelength, and phonons close to the BZ centre have comparatively long wavelengths. The second case is more interesting. It is clear that as one approaches $\bm{q}=2/3\langle111\rangle_{\mbox{\scriptsize p}}$, from either direction, there is a significant decrease in measured LA linewidths towards the predicted values, until, around the $2/3$ position they are coincident. We observe no change in TA linewidths over the similar region. This suggests that the linewidth, and hence lifetime, of phonons in the $\mbox{LA-}2/3\langle111\rangle_{\mbox{\scriptsize p}}$ mode, and to a lesser extent those nearby, are unaffected by the presence of the correlated disorder. The reason for the minimum at this position is currently unclear; however, it does indicate that this could be an allowed mode in both the global and local representations. Further inelastic scattering studies are required around this position, ideally with greater $\bm{q}\mbox{-resolution}$, to determine the true minimum and allow greater insight.

The phonon energies show good overall agreement with both theory and earlier experiments\cite{LLNL}, although the model is consistently underpredictive, likely due to the complexity involved in modelling $5f$ systems. Concerning detailed comparison between theory and experiment we highlight two points. Firstly, the observed $\mbox{LA-}2/3\langle111\rangle_{\mbox{\scriptsize p}}$ mode is hardened compared to theoretical predictions. The softening of this mode is a geometric effect present to varying extents in all monatomic \textit{bcc} crystals\cite{LA_2/3}. The mode corresponds to $\left[111\right]_{\mbox{\scriptsize p}}$ atomic chains performing a shearing motion that preserves interatomic distance, thus producing zero restoring force. The extent of the softening is determined by the degree of interchain forces present\cite{Ho}. We propose the local correlations increase interchain forces as sequential atoms along $\left[111\right]_{\mbox{\scriptsize p}}$ distort in antiphase, puckering the chains. The subsequent hardening has important consequences, especially when considering $\gamma\mbox{-UMo}$ as a potential advanced nuclear fuel\cite{U-Mo1,U-Mo2}, since this mode determines the activation enthalpy for self-diffusion\cite{LA_2/3} and understanding the diffusion characteristics of a nuclear fuel is important for accurately modelling the behaviour under operating conditions; this is discussed further below. Secondly, we observe a loss of longitudinal-transverse degeneracy at \textbf{P}. Degeneracy is imposed by \textit{bcc} symmetry\cite{bcc_degen} however high-symmetry positions from the parent BZ do not necessarily map onto high-symmetry positions in the superstructure BZ. As such, no degeneracy is required at \textbf{P}, leaving a direct fingerprint of the local symmetry reduction. Similar effects have been predicted for correlated compositional disorder, where one observes extra dispersive behaviour after correlations are introduced\cite{Aperiodic}. We also observe a moderate softening of the $\mbox{TA}\left[001\right]_{\mbox{\scriptsize p}}$ mode near the zone boundary, however we do not consider this to be a consequence of disorder-phonon coupling, further discussion is provided in supplemental material VII\cite{SM}.

\begin{figure}[t]
	\centering
	\includegraphics*[width=\linewidth]{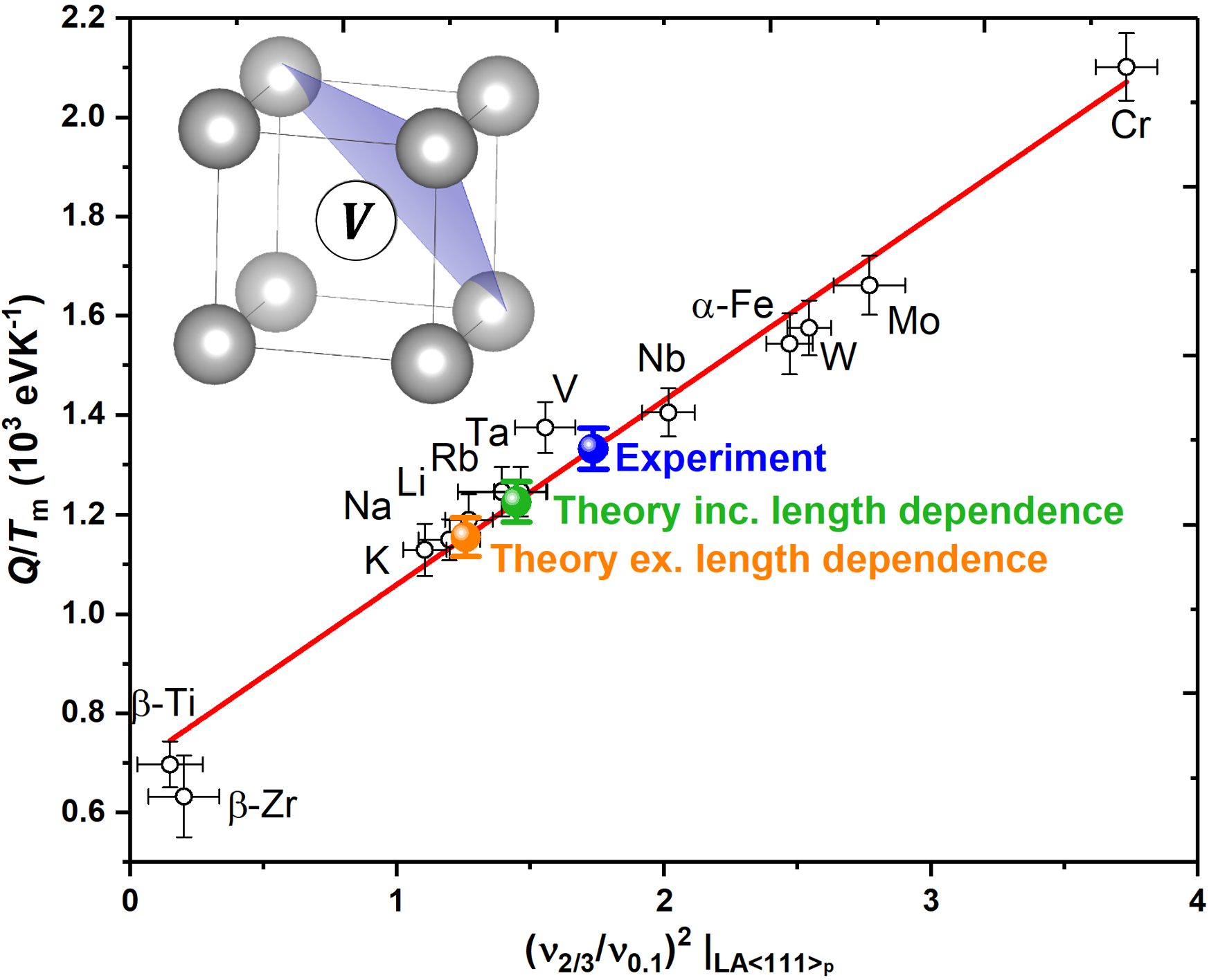}
	\caption{\textbf{ Relationship between the activation enthalpy for self-diffusion, \textit{Q}, in \textit{bcc} metals and the square of the $\bm{\mbox{LA-}2/3\langle111\rangle_{\mbox{\scriptsize p}}}$ frequency, $\bm{\left(\nu_{2/3}\right)^{2}}$, adapted from K\"ohler U. \& Herzig C.}\cite{LA_2/3} Universal scaling is achieved by normalising \textit{Q} to the melting temperature $\mbox{\textit{T}}_{\mbox{\scriptsize m}}$ and $\nu_{2/3}$ to $\nu_{0.1}$, the latter adjusting for lattice stiffness. Squared frequency ratios for both our experimental results and different theoretical approaches are shown as coloured spheres. These data sit on the universal curve by construction and errors are propagated from uncertainty in the linear fit. (Insert) Schematic \textit{bcc} unit cell with a central monovacancy, a plane containing a hopping saddle point is highlighted in blue.} 
	\label{Fig7}
\end{figure} 

Although the present theory\cite{Castellano} does not include correlated displacive disorder, we can obtain some indication of the effect random displacive disorder has on the $\mbox{LA-}2/3\langle111\rangle_{\mbox{\scriptsize p}}$ mode. Random disorder is likely to only affect the dispersion by modulating the IFC’s through a random distribution of bond lengths. Therefore, by allowing the IFC’s to be dependent on this parameter, as determined from the fully relaxed \textit{ab initio} molecular dynamics (AIMD) step, we can approximate this form of static disorder. To do this, we follow the idea of the bond-stiffness vs. bond-length model of Van de Walle \& Ceder\cite{VdW} and introduce a linear dependence of the IFC parameters as a function of the bond length
\begin{equation}
a\left(l\right) = a_{\scriptsize 0} + a_{\scriptsize 1}\left(l-l_{\scriptsize 0}\right),
\end{equation}
where $l$ and $l_{\scriptsize 0}$ are the length of the bond in the alloy and the ideal structure, respectively. $a\left(l\right)$ is the IFC coefficient, $a_{\scriptsize 0}$ is the stiffness of the bond at its ideal length $l_{\scriptsize 0}$, and $a_{\scriptsize 1}$ is a coefficient relating the stiffness of the coefficient to the deviation of the bond length from the ideal structure. To obtain the length-dependent IFC, the coefficients $a_{\scriptsize 0}$ and $a_{\scriptsize 1}$ are fitted from the AIMD runs, performed using the normal temperature dependant method, see supplemental material V\cite{SM}. After including length dependency, both modes, TA and LA, show increases by roughly $1\:\mbox{meV}$ for modes that contain uranium, the Mo-Mo modes may also undergo a slight increase, see supplemental material VIII \cite{SM}. This is consistent with our assessment that displacive disorder is the main factor in hardening the $\mbox{LA-}2/3\langle111\rangle_{\mbox{\scriptsize p}}$ mode. The remaining discrepancy between theory and experimental results is attributed to the short-range correlations, which are not captured theoretically.

As mentioned above, the energy of this mode has important practical ramifications. It is well established that in \textit{bcc} metals the energy of this mode may be explicitly linked to the activation enthalpy of self-diffusion, which is dominated by monovacancies\cite{LA_2/3}. This may be seen intuitively by considering the case of a monovacancy occupying the body centered atomic position, as shown in Figure \ref{Fig7}. When the mode is sufficiently soft, such that the amplitude of oscillation exceeds the marked saddle point, the monovacancy may hop from its central position to the corner position. This form of directly phonon-mediated monovacancy self-diffusion accounts for the anomalously high diffusion characteristics in \textit{bcc} metals with significant softening at the $\mbox{LA-}2/3\langle111\rangle_{\mbox{\scriptsize p}}$ position.

Figure \ref{Fig7} shows that from the measured $23$  at.\% Mo  phonon dispersion one would expect significant phonon-mediated monovacancy self-diffusion. It is also clear that by including an approximation of random displacive disorder the model better replicates the experimental result, underlining the role random displacive disorder plays in hardening the $\mbox{LA-}2/3\langle111\rangle_{\mbox{\scriptsize p}}$. However, there still exists a substantial discrepancy that can likely be attributed to short-range correlations. This highlights the need to develop theoretical tools that can accurately capture the effects of correlated disorder.

\subsection{Discussion on Disorder-Phonon Coupling}
Nanoscale complexity, like that observed in this study, has been shown to be of great importance in phonon engineering for emerging thermoelectrics\cite{complex}, efficiently scattering mid-long wavelength phonons that are responsible for the majority of phonon mediated heat transport in alloys\cite{mid-long}, and possibly contributing to the destruction of long-wavelength phonon coherence\cite{coherence}. Spontaneous structural modulations have been observed in numerous promising candidates\cite{AgSbTe2,ZnSb,PbTe,AgBiS2,PbS,Bozin,AgBiS2_2}. These systems are vastly more complex than the binary alloy investigated in our study, often involving chemical ordering in conjunction with structural distortions, however we observe similarly extraordinary phonon broadening, in both magnitude and the proportion of the dispersion affected. This indicates that the power of nanoscale structural modulations to suppress phonon lifetimes is not system specific, and instead, appears to be a generalised phenomena.

High entropy alloys are an alternate strategy that have garnered  interest as potential thermoelectrics and thermal barrier materials due to the maximal levels of chemical disorder present and the phonon lifetime suppression this entails\cite{HEA1,HEA2}. However, we show that, not only does the efficacy of nanoscale structural modulation vastly exceed the alloying effects present in our system, even considering the large mass difference between uranium and molybdenum\cite{Cu-Au}, it also exceeds those predicted for high entropy alloy systems\cite{HEA2}. This suggests that basis-lattice symettry mismatch and the resulting structurally degenerate groundstates may be a key ingredient for overcoming the limit set by maximal chemical disorder and designing systems with strongly suppressed lattice thermal conductivity. We also expect that the intrinsic tunability imbued in the disorder will be reflected in phonon lifetimes such that the degree of broadening will be related to $|\delta|$ and $\xi$. However, we do note that in a metallic system, like the alloys investigated in this study, the majority of heat will be carried by the free electrons such that the total thermal conductivity is dominated by its electronic component and remains significant even with a vanishingly small phononic component. To employ the disorder-phonon coupling demonstrated in this paper, and efficiently reduce total thermal conductivity, one would need to realise a similar level of correlated disorder in a semiconducting or insulating system; an example would be the thermoelectric system $\mbox{Ag}\mbox{Sb}\mbox{Te}_{2}$\cite{AgSbTe2,AgSbTe2_2}.

\section{Conclusions}
This work shows a new form of intrinsically tuneable correlated disorder that arises from the conflict created by a mismatch in preferred symmetry between a crystallographic basis and the lattice upon which it is arranged. We present a unique solution for the short-range superstructure given by a frozen $\mbox{TA}_1$ phonon, simultaneously recovering $Cmcm$ symmetry, reminiscent of $\alpha\mbox{-uranium}$, as well as providing a natural explanation for the prolate ellipsoidal diffuse signal. Furthermore, by combining grazing incidence inelastic x-ray scattering and state-of-the-art \textit{ab initio} molecular dynamics simulations we discover strong disorder-phonon coupling. This dramatically suppresses phonon-lifetimes compared to alloying alone, hardens the $\mbox{LA-}2/3\langle111\rangle_{\mbox{\scriptsize p}}$ mode ubiquitous to monotonic \textit{bcc} crystals, and relaxes degeneracy conditions at the \textbf{P} position.

These studies highlight basis-lattice symmetry mismatch and the resulting correlated disorder as possibly a key ingredient for future phonon engineering strategies that aim to efficiently suppress lattice thermal conductivity in thermoelectric or thermal barrier materials. Potentially also allowing the lower limit set by maximal chemical disorder to be overcome. Conversely, local conditions must be considered when designing materials for heat transfer applications were high thermal conductivity is desired. The effects are also expected to extend to other periodic phenomena, electronic structure, spin waves etc. and in systems supportive of greater $|\delta|$ could have significant impact on properties that are strongly dependent on interatomic distance. Of course, these effects are not limited to alloys or compounds containing the intermediate actinides, any element with a low symmetry groundstate is a promising candidate, technologically vital examples would be P, S, Bi and Ga. We believe this new form of correlated disorder, with the strength of the effects and intrinsic tuneability imbued in the alloy composition, constitutes a compelling tool for designing disorder into functional materials.

\begin{acknowledgements}
We thank C. Howard for discussions and advice. We also thank AWE for providing the depleted uranium sputtering target used in this work. D. Chaney acknowledges combined funding from the ESPRC and the AWE through the I-CASE program. D. Chaney also acknowledges funding from the ESRF through their traineeship program. Finally we note the following copyright: UK Ministry of Defence \textcopyright\space Crown Owned Copyright 2020/AWE
\end{acknowledgements}

\clearpage


\begin{thebibliography}{5}

\bibitem{Bragg}W. L. Bragg, \textit{The diffraction of short electromagnetic waves by a crystal}, Proc. Cambridge Phil. Soc. \textbf{17,} 43-57 (1913).
\bibitem{100years}T. R. Welberry,  \&  T. Weber,  \textit{One hundred years of diffuse scattering}, Crystallography Reviews \textbf{22,} 2-78 (2016).
\bibitem{Diffuse Scattering}R. I. Barabash, G. E. Ice,  \& P. E. A. Turchi,  \textit{Diffuse Scattering and the Fundamental Properties of Materials}, (Momentum Press, New York, 2009).
\bibitem{Nanoscale_Structure}S. J. L. Billinge, \& I. Levin, \textit{The problem with determining atomic structure at the nanoscale}, Science \textbf{316,} 561-565 (2007).
\bibitem{Goodwin1}D. A. Keen,  \& A. L. Goodwin,  \textit{The crystallography of correlated disorder}, Nature \textbf{521,} 303-309 (2015).
\bibitem{Goodwin2} A. Simonov, \& A. L. Goodwin, \url{https://arxiv.org/abs/1912.00366}.
\bibitem{Aperiodic}A. R. Overy,  A. B. Cairns, M. J. Cliffe, A. Simonov, M. G. Tucker, \& A. L. Goodwin, \textit{Design of crystal-like aperiodic solids with selective disorder-phonon coupling}, Nature Communications \textbf{7,} 10445 (2016).
\bibitem{ferro1}M. S. Senn, D. A. Keen,  T. C. A. Lucas, J. A. Hriljac, \& A. L. Goodwin, \textit{Emergence of long-range order in $\mbox{BaTiO}_3$ from local symmetry-breaking distortions}, Phys. Rev. Lett. \textbf{116,} 207602 (2016)
\bibitem{ferro2}G. Perversi, E. Pachoud, J. Cumby, J. M. Hudspeth, J. P. Wright, S. A. J. Kimber, \& J. P. Attfield, \textit{Co-emergence of magnetic order and structural fluctuations in magnetite}, Nature Communications \textbf{10,} 2857 (2019).
\bibitem{ferro3}M. J. Krogstad, P. M. Gehring, S. Rosenkranz, R. Osborn, F. Ye, Y. Liu, J. P. C. Ruff, W. Chen, J. M. Wozniak, H. Luo \textit{et al.}, \textit{The relation of local order to material properties in relaxor ferroelectrics}, Nature Materials \textbf{17,} 718-724 (2018).
\bibitem{Bozin}E. S. Bo\v{z}in, C. D. Malliakas, P. Souvatzis, T. Proffen, N. A. Spaldin, M. G. Kanatzidis, \& S. J. L. Billinge, \textit{Entropically stabilized local dipole formation in lead chalcogenides}, Science \textbf{330,} 1660-1663 (2010).
\bibitem{AgSbTe2}J. Ma, O. Delaire, A. F. May, C. E. Carlton, M. A. McGuire, L. H. VanBebber, D. L. Abernathy, G. Ehlers, T. Hong, A Huq \textit{et al.}, \textit{Glass-like phonon scattering from a spontaneous nanostructure in $\mbox{AgSbTe}_2$}, Nature Nanotechnology \textbf{8,} 445-451 (2013).
\bibitem{PbTe}B. Sangiorgio, E. S. Bo\v{z}in, C. D. Malliakas, M. Fechner, A. Simonov, M. G. Kanatzidis, S. J. L. Billinge, N. A. Spaldin, \& T. Weber, \textit{Correlated local dipoles in PbTe}, Phys. Rev. Mater. \textbf{2,} 085402 (2018).
\bibitem{photovoltaic}M. T. Weller, O. J. Weber, P. F. Henry, A. M. Di Pumpo,  \& T. C. Hansen, \textit{Complete structure and cation orientation in the perovskite photovoltaic methylammonium lead iodine between 100 and 325 K.} Chemical Communications \textbf{51,} 4180-4183 (2015).
\bibitem{ionic1}A. D\"uvel, P. Heitjans, P. Fedorov, G. Scholz, G. Cibin, A. V. Chadwick, D. M. Pickup, S. Ramos, L. W. L. Sayle, E. K. L. Sayle \textit{et al.}, \textit{Is geometric frustration-induced disorder a recipe for high ionic conductivity?}, J. Am. Chem. Soc. \textbf{139,} 5842-5848 (2017).
\bibitem{functional}A. L. Goodwin, \textit{Opportunities and challenges in understanding complex functional materials}, Nature Communications \textbf{10,} 4461 (2019).
\bibitem{Mettout}B. Mettout, V. P. Dmitriev, M. Ben Jaber, \& P. Tol\'edano, \textit{Theory of reconstructive transformations in actinide elements: packing of nonspherical atoms and macroscopic symmetries}, Phys. Rev. B \textbf{48,} 6908 (1993).
\bibitem{Soderlind}P. S\"oderlind, O. Eriksson, B. Johansson, J. M. Wills, \& A. M. Boring, \textit{A unified picture of the crystal structures of metals}, Nature \textbf{374,} 524-525 (1995).
\bibitem{Historical Persepctive}G. H. Lander, E. S. Fisher, \& S. D. Bader, \textit{The solid-state properties of uranium a historical perspective and review}, Advances in Physics \textbf{43,} 1-111 (1994).
\bibitem{Yakel}H. L. Yakel, \textit{Crystal structures of transition phases formed in $\mbox{U/}16.60$ at\% $\mbox{Nb/}5.64$ at\% Zr alloys}, Journal of Nuclear Materials \textbf{33,} 286-295 (1969). 
\bibitem{LLNL}Z. E. Brubaker, S. Ran, A. H. Said, M. E. Manley, P. S\"{o}derlind, D. Rosas, Y. Idell, R. J. Zieve, N. P. Butch, \& J. R. Jeffries, \textit{Phonon dispersion of Mo-stabilized $\gamma\mbox{-U}$ measured using inelastic x-ray scattering}, Phys. Rev. B \textbf{100,} 094311 (2019).
\bibitem{Monolithic}R. M. Hengstler, L. Beck, H. Breitkreutz, C. Jarousse, R. Jungwirth, W. Petry, W. Schmid, J. Schneider, \& N. Wieschalla, \textit{Physical properties of monolithic U8 wt.\%-Mo}, Journal of Nuclear Materials \textbf{402,} 74-80 (2010).
\bibitem{Lopes}D. A. Lopes, T. A. G. Restivo, \& A. F. Padilha, \textit{Mechanical and thermal behaviour of U-Mo and U-Nb-Zr alloys}, Journal of Nuclear Materials \textbf{440,} 304-309 (2013).
\bibitem{Total}L. R. Owen, H. Y. Playford, H. J. Stone, \& M. G. Tucker, \textit{A new approach to the analysis of short-range order in alloys using total scattering}, Acta Materialia \textbf{115,} 155-166 (2016).
\bibitem{Adamska}A. M. Adamska, R. Springell, \& T. B. Scott, \textit{Characterization of poly- and single-crystal uranium-molybdenum alloy thin films}, Thin Solid Films \textbf{550,} 319-325 (2014).
\bibitem{Dwight}A. E. Dwight, \textit{The uranium-molybdenum equilibrium diagram below $900\degree\mbox{C}$}, Journal of Nuclear Materials \textbf{1,} 81-87 (1960).
\bibitem{match}R. C. C. Ward, E. J. Grier, \& A. K. Petford-Long, \textit{MBE growth of $(110)$ refractory metals on \textit{a}-plane sapphire}, Journal of Materials Science: Materials in Electronics \textbf{14,} 533-539 (2003).
\bibitem{IXS}M. Krisch, \& F. Sette, \textit{Light Scattering in Solids IX pp 317-370} (Springer, Berlin, 2006).
\bibitem{Diffuse}A. Girard, T. Nguyen-Thanh, S. M. Souliou, M. Stekiel, W. Morgenroth, L. Paolasini, A. Minelli, D. Gambetti, B. Winkler \& A. Bosak, \textit{A new diffractometer for diffuse scattering studies on the ID28 beamline at the ESRF}, J. Synchrotron Rad. \textbf{26,} 272-279 (2019).
\bibitem{Crysalis}Agilent (2014). \textit{CrysAlis Pro}. Agilent Technologies Ltd, Yarnton, Oxfordshire, England. 
\bibitem{Rennie}S. Rennie, E. Lawrence Bright, J. E. Darnbrough, L. Paolasini, A. Bosak, A. D. Smith, N. Mason, G. H. Lander, \& R. Springell, \textit{Study of phonons in irradiated epitaxial thin films of $\mbox{UO}_{2}$}, Phys. Rev. B \textbf{97,} 224303 (2018).
\bibitem{ABINIT}X. Gonze, B. Amadon, G. Antonius, F. Arnardi, L. Baguet, J. -M. Beuken, J. Bieder, F. Bottin, J. Bouchet, E. Bousquet \textit{et al.}, \textit{The ABINIT project: impact, environment and recent developments}, Comput. Phys. Commun. \textbf{248,} 107042 (2020). 
\bibitem{Castellano}A. Castellano, F. Bottin, B. Dorado, \& J. Bouchet,  \textit{Thermodynamic stabilization of $\gamma\mbox{-U---Mo}$: effects of Mo content and temperature}, Phys. Rev. B \textbf{101,} 184111 (2020).
\bibitem{SM}See Supplemental Material at [url] for more details on the parent-superstructure relationship, details of theoretical modelling and further, subsidiary analysis.
\bibitem{U2Mo}E. K. Halteman, \textit{The crystal structure of $\mbox{U}_2\mbox{Mo}$}, Acta Cryst. \textbf{10,} 166-169 (1957).
\bibitem{delta-Ti}Y. Akahama, H. Kawamura, \& T. L. Bihan, \textit{New $\delta\:\left(\mbox{Distorted-\textit{bcc}}\right)$ titanium to $220\:\mbox{GPa}$} Phys. Rev. Lett. \textbf{87,} 275503 (2001).
\bibitem{Bosak}A. A. Bosak, S. B. Vakhrushev, A. A. Naberezhnov, \& P. Y. Vanina,  \textit{Peculiarities of diffuse synchrotron radiation scattering in the SBN-60 single crystal at room temperature}, St Petersburg Polytechnical Univ. J. Phys. Math. \textbf{1,} 235-238 (2015). 
\bibitem{Johann}J. Bouchet, \& F. Bottin, \textit{High-temperature and high-pressure phase transitions in uranium}, Phys. Rev. B \textbf{95,} 054113 (2017).
\bibitem{Soderlind2}P. S\"oderlind, B. Grabowski, L. Yang, A. Landa, T. Bj\"{o}rkman, P. Souvatzis, \& O. Eriksson, \textit{High-temperature phonon stabilization of $\gamma\mbox{-uranium}$ from relativistic first-principles theory}, Phys. Rev. B \textbf{85,} 060301 (2012).
\bibitem{Beta}A. C. Lawson, C. E. Olsen, J. W. Richardson Jnr, M. H. Mueller, \& G. H. Lander, \textit{Structure of $\beta\mbox{-uranium}$}, Acta Cryst. B \textbf{44,}  89-86 (1988).
\bibitem{Axe}J. D. Axe, G. Gr\"ubel, \& G. H. Lander, \textit{Structure and phase transformations in uranium metal}, Journal of Alloys and Compounds \textbf{213/214,} 262-267 (1994).
\bibitem{GRIX}J. Serrano, A. Bosak, M. Krisch, F. J. Manj\'{o}n, A. H. Romero, N. Garro, X. Wang, A. Yoshikawa, \& M. Kuball, \textit{InN thin film lattice dynamics by grazing incidence inelastic x-ray scattering}, Phys. Rev. Lett. \textbf{106,} 205501 (2011).
\bibitem{omega}J. D. Axe, D. T. Keating, \& S. C. Moss, \textit{Anomalous inelastic neutron scattering in bcc Zr-Nb alloys}, Phys. Rev. Lett. \textbf{35,} 531 (1975).
\bibitem{omega2}Y. Yamada, \& K. Fuchizaki,  \textit{Anomalous lattice-dynamical properties of a quenched diffuse $\omega$ phase in Zr-Nb alloys}, Phys. Rev. B \textbf{42,} 9240 (1990).
\bibitem{LA_2/3}U. K\"olher, \& C. Herzig,  \textit{On the correlation between self-diffusion and the low-frequency $\mbox{LA-}\nicefrac{2}{3}\langle111\rangle$ phonon mode in b.c.c. metals}, Philos. Mag. A \textbf{58,} 769-786 (1998).
\bibitem{Ho}K. -M. Ho, C. L. Fu, \& B. N. Harmon, \textit{Microscopic analysis of interatomic forces in transition metals with lattice distortions}, Phys. Rev. B \textbf{28,} 6687 (1983).
\bibitem{U-Mo1}J. Rest, Y. S. Kim, G. L. Hofman, M. K. Meyer, \& S. L. Hayes, \textit{U-Mo Fuels Handbook Version $1.0$} (Argonne National Laboratory, Argonne, 2006).
\bibitem{U-Mo2}S. Van den Berghe, \& P. Lemoine, \textit{Review of 15 years of high-density low-enriched UMo dispersion fuel development for research reactors in Europe}, Nucl. Eng. Technol. \textbf{46,} 125-146 (2014).
\bibitem{bcc_degen}Y. Chen, J. Ma, S. Wen, \& W. Li, \textit{Body-centered-cubic structure and weak anharmonic phonon scattering in tungsten}, npj Comput. Mater. \textbf{5,} 98 (2019).
\bibitem{VdW}A. Van de Walle, \& G. Ceder, \textit{The effect of lattice vibrations on substitutional alloy thermodynamics}, Rev. Mod. Phys. \textbf{74,} 11 (2002).
\bibitem{complex}G. J. Snyder, \& E. S. Toberer,  \textit{Complex thermoelectric materials}, Nature Materials \textbf{7,} 105-114 (2008).
\bibitem{mid-long}W. Kim, J. Zide, A. Gossard, D. Klenov, S. Stemmer, A. Shakouri, \& A. Majumdar, \textit{Thermal conductivity reduction and thermoelectric figure of merit increase by embedding nanoparticles in crystalline semiconductors}, Phys. Rev. Lett. \textbf{96,} 045901 (2006) 
\bibitem{coherence}M. N. Luckyanova, J. Garg, K. Esfarjani, A. Jandl, M. T. Bulsara, A. J. Schmidt, A. J. Minnich, S. Chen, M. S. Dresselhaus, Z. Ren \textit{et al.}, \textit{Coherent phonon heat conduction in superlattices}, Science \textbf{338,} 936-939 (2012).
\bibitem{ZnSb}H. J. Kim, E. S. Bo\v{z}in, S. M. Haile, G. J. Synder, \& S. J. L. Billinge, \textit{Nanoscale $\alpha$-structural domains in phonon-glass thermoelectric material $\beta\mbox{-Zn}_4\mbox{Sb}_3$}, Phys. Rev. B \textbf{75,} 134103 (2007). 
\bibitem{AgBiS2}E. Rathmore, R. Juneja, S. P. Culver, N. Minafra, A. K. Singh, W. G. Zeier, \& K. Biswas, \textit{Origin of ultralow thermal conductivity in n-type cubic bulk $\mbox{AgBiS}_2$: soft Ag vibrations and local structural distortion induced by the Bi $6s^2$ lone pair}, Chem. Mater. \textbf{6,} 2106-2113 (2019).
\bibitem{AgBiS2_2}J. L. Niedziela, D. Bansal, J. Ding, T. Lanigan-Atkins, C. Li, A. F. May, H. Wang, J. Y. Y. Lin, D. L. Abernathy, G. Ehlers, \textit{et al}, \textit{Controlling phonon lifetimes via sublattice disordering in $\mbox{AgBiS}_2$}, Phys. Rev. Materials, \textbf{4,} 105402 (2020).
\bibitem{PbS}C. M. Zeuthen, P. S. Thorup, N. Roth, \& B. B. Iversen, \textit{Reconciling crystallographic and physical property measurements on thermoelectric lead sulfide}, J. Am. Chem. Soc. \textbf{141,} 8146-8157 (2019).
\bibitem{HEA1}Z. Fan, H. Wang, Y. Wu, X. J. Liu, \& Z. P. Lu, \textit{Thermoelectric high-entropy alloys with low lattice thermal conductivity}, RSC Adv. \textbf{6,} 52164 (2016).
\bibitem{HEA2}F. K\"{o}rmann, Y. Ikeda, B. Grabowski, \& M. H. F. Sluiter, \textit{Phonon broadening in high entropy alloys}, npj Comput. Mater. \textbf{3,} 36 (2017).
\bibitem{Cu-Au}Y. Ikeda, A. Carreras, A. Seko, A. Togo, \& I. Tanaka, \textit{Mode decomposition based on crystallographic symmetry in the band-unfolding method}, Phys. Rev. B \textbf{95,} 024305 (2017).



\textbf{Below References Appear in Supplemental Material Only}\newline
\bibitem{chemicalorder1}S. V. Strelova, Y. S. Umansky, \& O. S. Ivanov,  \textit{Short-range order in uranium-niobium solid solution}, J. Nucl. Mater. \textbf{34,} 160-164 (1970).
\bibitem{chemicalorder2}K. Tangri, \textit{Les phases gamma m\'etastables dans les alliages d'uranium contenant du molybdn\`ene}, Mem. Sci. Rev. Met. \textbf{58,} 469-477 (1961).

\bibitem{FWHM}O. Young, A. R. Wildes, P. Manuel, B. Ouladdiaf, D. D. Khalyavin, G. Balakrishnan, \& O. A. Petrenko, \textit{Highly frustrated magnetism in $\mbox{SrHo}_2\mbox{O}_4$: coexistence of two types of short-range order}, Phys. Rev. B \textbf{88,} 024411 (2013). 

\bibitem{Band_strucutre}V. Popescu, \& A. Zunger,  \textit{Effective band structure of random alloys}, Phys. Rev. Lett. \textbf{104,} 236403 (2010).
\bibitem{Starikov}S. V. Starikov, L. N. Kolotova, A. Y. Kuksin, D. E. Smirnova, \& V. I. Tseplyaev, \textit{Atomistic simulation of cubic and tetragonal phases of U-Mo alloy: structure and thermodynamic properties}, J. Nucl. Mater. \textbf{499,} 451-463 (2018).
\bibitem{SQRS}A. Zunger, S.-H. Wei, L. G. Ferreira, \& J. E. Bernard, \textit{Special quasirandom structures}, Phys. Rev. Lett. \textbf{65,} 353 (1990).
\bibitem{SQS2}A. Van de Walle, P. Tiwary, M. de Jong, D. L. Olmsted, M. Asta, A. Dick, D. Shin, Y. Wang, L. -Q. Chen, \& Z. -K. Liu, \textit{Efficient stochastic generation of special quasirandom structures}, Calphad \textbf{42,} 13-18 (2013).
\bibitem{TDEP}O. Hellman, P. Steneteg, I. A. Abrikosov, \& S. I. Simak, \textit{Temperature dependant effective potential method for accurate free energy calculations of solids}, Phys. Rev. B \textbf{87,} 104111 (2013).
\bibitem{HT-HP}J. Bouchet, \& F. Bottin,  \textit{High-temperature and high-pressure transitions in uranium}, Phys. Rev. B \textbf{95,} 054113 (2017).
\bibitem{A-TDEP}F. Bottin, J. Biedar, \& J. Bouchet, \textit{A-TDEP: temperature dependant effective potential for ABINIT - lattice dynamic properties including anharmonicity}. Comput. Phys. Commun. \textbf{254,} 107301 (2020). 
\bibitem{Shulumba}N. Shulumba, O. Hellman, Z. Raza, B. Alling, J. Barrirero, F. M\"{u}cklich, I. A. Abrikosov, \& M. Od\'{e}n, \textit{Lattice vibrations change the solid solubility of an alloy at high temperatures}, Phys. Rev. Lett. \textbf{117,} 205502 (2016). 
\bibitem{BZ_lifetimes}A. Glensk, B. Grabowski, T. Hickel, J. Neugebauer, J. Neuhaus, K. Hradil, W. Petry, \& M. Leitner, \textit{Phonon lifetimes throughout the brillouin zone at elevated temperatures from experiment and ab initio}, Phys. Rev. Lett. \textbf{123,} 235501 (2019).
\bibitem{Pang}J. W. L. Pang, W. J. L. Buyers, A. Chernatynskiy, M. D. Lumsden, B. C. Larson, \& S. R. Phillpot, \textit{Phonon lifetime investigation of anharmonicity and thermal conductivity of $\mbox{UO}_2$ by neutron scattering and theory}, Phys. Rev. Lett. \textbf{110,} 157401 (2013).
\bibitem{AgSbTe2_2}J. Ma, O. Delaire, E. D. Specht, A. F. May, O. Gourdon, J. D. Budai, M. A. McGuire, T. Hong, D. L. Abernathy, G. Ehlers, \& E. Karapetrova, \textit{Phonon scattering rates and atomic ordering in $\mbox{Ag}_{1-x}\mbox{Sb}_{1+x}\mbox{Te}_{2+x}\:\left(x=0,0.1,0.2\right)$ investigated with inelastic neutron scattering and synchrotron diffraction}, Phys. Rev. B \textbf{90,} 134303 (2014).





\end{thebibliography}
\end{document}


\setlength{\abovedisplayskip}{10pt}
\setlength{\belowdisplayskip}{10pt}

\title{\LARGE Supplemental Material for ``Tuneable Correlated Disorder in Alloys"}
\author{D. Chaney}
\email{daniel.chaney@bristol.ac.uk}
\affiliation{School of Physics, University of Bristol, Tyndall Avenue, Bristol, BS8 1TL, UK}
\affiliation{European Synchrotron Radiation Facility, BP 220, F-38043 Grenoble, France}
\author{A. Castellano}
\affiliation{CEA, DAM, DIF, F-91297 Arpajon, France}
\author{A. Bosak}
\affiliation{European Synchrotron Radiation Facility, BP 220, F-38043 Grenoble, France}
\author{J. Bouchet}
\author{F. Bottin}
\author{B. Dorado}
\affiliation{CEA, DAM, DIF, F-91297 Arpajon, France}
\author{L. Paolasini}
\affiliation{European Synchrotron Radiation Facility, BP 220, F-38043 Grenoble, France}
\author{S. Rennie}
\author{C. Bell}
\author{R. Springell}
\email{phrss@bristol.ac.uk}
\author{G. H. Lander}
\affiliation{School of Physics, University of Bristol, Tyndall Avenue, Bristol, BS8 1TL, UK}

\date{\today}

\begin{center}
{UK Ministry of Defence \textcopyright\space Crown Owned Copyright 2020/AWE}
\end{center} 

\maketitle
\thispagestyle{plain}

\section{Diffuse reflections from additional superstructure domains}
\begin{figure}[h]
	\centering
	\includegraphics*[width=\linewidth]{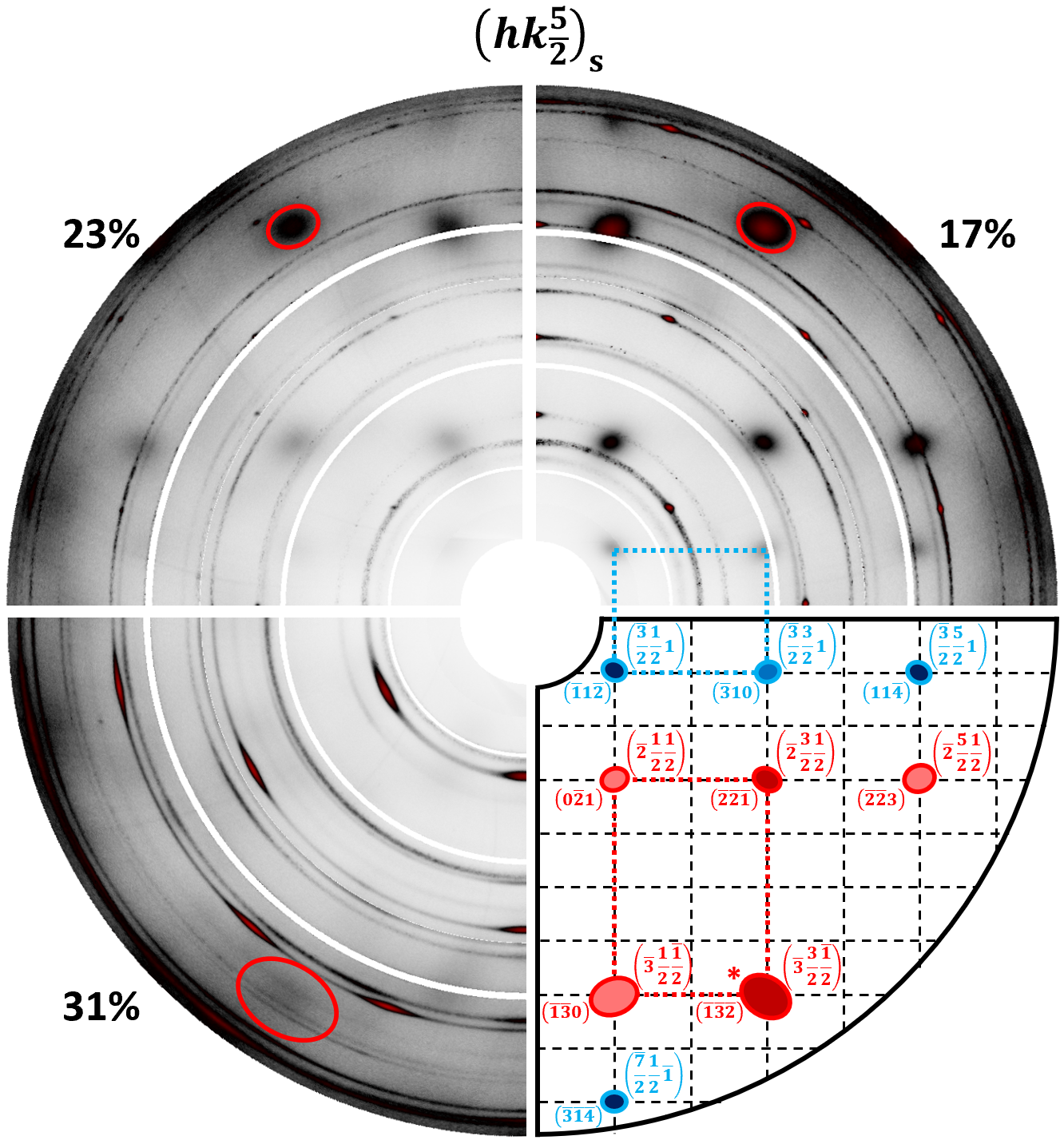}
	\caption{\textbf{ Composite reciprocal space reconstruction of the $\mathbf{\left(hk \frac{5}{2}\right)_{\mbox{\scriptsize s}}}$ plane for different Mo content.} The notation used is the same as for Figure $2$ - main text. Red open ellipses overlaid on the data show equivalent points in each quadrant with the corresponding reflection marked by an asterisk in the schematic (bottom right). In the schematic, domains with $\bm{b}_{\mbox{\scriptsize s}}\equiv \bm{a}_{\mbox{\scriptsize p}}$ are shown in red, and those with $\bm{b}_{\mbox{\scriptsize s}}\equiv \bm{c}_{\mbox{\scriptsize p}}$  in blue. Non-integer indices are from one possible \textit{bcc} representation, while integer indices correspond to the relevant choice of \textit{Cmcm} domain. Lighter (darker) red and blue shading corresponds to clockwise (anti-clockwise) domains, respectively.} 
	\label{FigS1}
\end{figure}

As explained in the main paper the \textit{Cmcm} superstructure produces $12$ diffuse reflections, from $6$ domains, one at every \textbf{N} position on the \textit{bcc} Brillouin zone (BZ) boundary. In Figure \textcolor{blue}{S}\ref{FigS1} we show the $4$ domains not shown in Figure $2$ - main text, namely those with $\bm{b}_{\mbox{\scriptsize s}}$ equivalent to either $\bm{a}_{\mbox{\scriptsize p}}$ (red) or $\bm{c}_{\mbox{\scriptsize p}}$ (blue). It is also possible to distinguish between reflections that originate from clockwise or anti-clockwise rotations about $\bm{b}_{\mbox{\scriptsize s}}$ as only one choice satisfies the condition $\left(h_{\mbox{\scriptsize s}}+k_{\mbox{\scriptsize s}}\right) = 2n$ imposed by a \textit{C}-face centring (where $n$ is an integer). A further condition within the \textit{Cmcm} structure is that reflections described by $\left(k_{\mbox{\scriptsize s}}+l_{\mbox{\scriptsize s}}\right) = 2n$ coincide with \textit{bcc} Bragg positions whereas diffuse peaks are described by $\left(k_{\mbox{\scriptsize s}}+l_{\mbox{\scriptsize s}}\right) = 2n+1$. The systematic coincidence of \textit{Cmcm} with \textit{bcc} positions shows the local superstructure is coherent with the global structure. In combination with Figure $2$ - main text, it can now be seen that the positive relationship between intensity and $k_{\mbox{\scriptsize s}}$ is present for all $6$ domains. Hence reflections along the horizontal (fixed $k_{\mbox{\scriptsize s}}$) of Figure \textcolor{blue}{S}\ref{FigS1} show similar intensities whereas those along the vertical (increasing $k_{\mbox{\scriptsize s}}$) show a marked increase in intensity. It should be noted that the same orientation matrix, which relates the diffractometer motor positions to the unit cell orientation, was used to construct both Figure $2$ - main text and Figure \textcolor{blue}{S}\ref{FigS1}, and corresponds to the clockwise domain $\bm{b}_{\mbox{\scriptsize s}}\equiv \bm{b}_{\mbox{\scriptsize p}}$. Hence, there are no reflections (Bragg or diffuse) from the chosen domain in Figure \textcolor{blue}{S}\ref{FigS1}, as this corresponds to a half-integer plane in the chosen representation. 
\clearpage
\clearpage
\section{Modelling the diffuse intensity}
Direct comparison between the diffuse and Bragg intensities is difficult as the detector is saturated at Bragg positions. However, we can determine that the integrated diffuse intensity, as shown in Figure \textcolor{blue}{S}\ref{FigS2}  (Fig. 2 - main text) is less than $1\%$ of the relevant Bragg peak. Using a frozen phonon model, we make a rough calculation of expected intensities. This model accurately captures the systematic absences at $k_{\mbox{\scriptsize s}}=0$ and the positive relationship between intensity and $k_{\mbox{\scriptsize s}}$, with an approximate dependence of $I\propto k_{\mbox{\scriptsize s}}^2$ modified by the scattering factor. This contrasts with normal Bragg reflections, including those from $\mbox{U}_{2}\mbox{Mo}$ precipitates, where scattering decreases with total momentum transfer, $\bm{Q}$, due to the scattering factor. 

Table S1 gives calculated intensities for all observed reflections in Figure $2$ - main text, as well a selection of extra reflections for $\left|\delta\right|=0.01$. In the absence of a distortion, the \textit{bcc} Bragg reflections would have a structure factor of $4$, however the deviation for small $\left|\delta\right|$ is not detectable experimentally. It also must be recognized that this model takes no account for the finite nature of the correlations, instead treating the system as a distorted, but perfectly ordered crystal where all reflections are “Bragg-like”.  In practice, the total integrated intensities predicted here are spread out over the entire diffuse peak as shown in Figure \textcolor{blue}{S}\ref{FigS1}. As discussed in the main text the observation of equal intensity diffuse signal at all \textbf{N} points indicates the domains are equally distributed, as expected from the degeneracy in the relevant frozen phonon mode. With much thinner films, strain effects from the substrate/buffer might allow one to selectively reduce the number of domains. This model makes no attempt to account for chemical ordering, however, as stated in the main text we observe no sign of this, early work suggesting such ordering did not use single crystals\cite{chemicalorder1,chemicalorder2}.

\clearpage
\begin{table}[h]
\centering
\renewcommand{\arraystretch}{0.9}
	\begin{tabular}{ |c|c|c|c|c|c| } 
		\hline
		\textbf{\textit{Cmcm} $\bm{(hkl)_{\mbox{\scriptsize s}}}$} & \textbf{\textit{bcc} $\bm{(hkl)_{\mbox{\scriptsize p}}}$} & $\bm{\sin\theta/\lambda\:\left(\mbox{\AA}^{-1}\right)}$ & $\bm{\left|\mbox{SF}\right|}$ & $\bm{\mbox{f}\left(Q\right)}$ & $\bm{\mbox{I}\left(\mbox{norm}\right)}$ \\ 
		\hline
		$\bm{\bar{2}02}$ & $\bm{\bar{2}00}$ & $\bm{0.293}$ & $\bm{4.00}$ & $\bm{0.743}$ &  $\bm{778}$ \\
		$\bm{\bar{4}02}$ & $\bm{\bar{3}0\bar{1}}$ & $\bm{0.464}$ & $\bm{4.00}$ & $\bm{0.614}$ & $\bm{466}$ \\
		& & & & & \\
		$\bar{1}12$&$\bar{\frac{3}{2}}1\frac{1}{2}$  & $0.274$ & $0.251$ & $0.760$ & $0.54$ \\
		$\bar{3}12$ & $\bar{\frac{5}{2}}1\bar{\frac{1}{2}}$ & $0.402$ & $0.251$ & $0.655$ & $0.37$ \\
		& & & & & \\
		$\bm{022}$ & $\bm{\bar{1}21}$ & $\bm{0.359}$ & $\bm{3.97}$ & $\bm{0.687}$ & $\bm{628}$\\
		$\mathit{\;\;\;\;\;\;\;\;\;\;\:\bar{1}22\:\left(\bar{2}2\bar{1}\right)}$ & $\mathit{\bar{\frac{3}{2}}2\frac{1}{2}}$ & $\mathit{0.374}$ & $\mathit{0.501}$ & $\mathit{0.676}$ & $\textit{1.6}$\\
		$\bm{\bar{2}22}$ & $\bm{\bar{2}20}$ & $\bm{0.415}$ & $\bm{3.97}$ & $\bm{0.646}$ & $\bm{532}$\\
		$\mathit{\;\;\;\;\;\;\;\;\;\;\:\bar{3}22\:\left(\bar{2}2\bar{3}\right)}$ & $\mathit{\bar{\frac{5}{2}}2\bar{\frac{1}{2}}}$ & $\mathit{0.475}$ & $\mathit{0.501}$ & $\mathit{0.606}$ & $\textit{1.2}$ \\
		$\bm{\bar{4}22}$ & $\bm{\bar{3}2\bar{1}}$ & $\bm{0.549}$ & $\bm{3.97}$ & $\bm{0.563}$ & $\bm{355}$ \\
		& & & & & \\
		$\bar{1}32$ & $\bar{\frac{3}{2}}3\frac{1}{2}$ & $0.497$ & $0.749$ & $0.593$ & $2.5$ \\
		$\bar{3}32$ & $\bar{\frac{5}{2}}3\bar{\frac{1}{2}}$ & $0.577$ & $0.749$ & $0.547$ & $1.95$ \\
		\hline
		$\mathit{\;\;\;\;\;\;\;\;\;\;\:01\bar{1}\:\left(110\right)}$ & $\mathit{\frac{1}{2}1\bar{\frac{1}{2}}}$ & $\mathit{0.180}$ & $\mathit{0.251}$ & $\mathit{0.852}$ & $\textit{0.715}$ \\
		$110$ & $\frac{1}{2}1\frac{1}{2}$ & $0.180$ &  $0.251$ & $0.852$ & $0.715$ \\
		$\bm{11\bar{1}}$ & $\bm{110}$ & $\bm{0.207}$ & $\bm{3.992}$ & $\bm{0.824}$ & $\bm{1000}$ \\
		$11\bar{2}$ & $\frac{3}{2}1\bar{\frac{1}{2}}$ & $0.274$ & $0.251$ & $0.760$ & $0.54$ \\
		$\mathit{\;\;\;\;\;\;\;\;\;\;\:21\bar{1}\:\left(112\right)}$ & $\mathit{\frac{3}{2}1\frac{1}{2}}$ & $\mathit{0.274}$ & $\mathit{0.251}$ & $\mathit{0.760}$ & $\textit{0.54}$ \\
		& & & & & \\
		$\mathit{\;\;\;\;\;\;\;\;\;\;\:120\:\left(021\right)}$ & $\mathit{\frac{1}{2}2\frac{1}{2}}$ & $\mathit{0.311}$ & $\mathit{0.501}$ & $\mathit{0.727}$ & $\mathit{1.95}$ \\
		$22\bar{1}$ & $\frac{3}{2}2\frac{1}{2}$ & $0.374$ & $0.501$ & $0.676$ & $1.65$ \\
		$\bm{220}$ & $\bm{121}$ & $\bm{0.359}$ & $\bm{3.97}$ & $\bm{0.687}$ & $\bm{628}$ \\
		$221$ & $\frac{1}{2}2\frac{3}{2}$ & $0.374$ & $0.501$ & $0.676$ & $1.65$ \\
		$\mathit{\;\;\;\;\;\;\;\;\;\;\:320\:\left(023\right)}$ & $\mathit{\frac{3}{2}2\frac{3}{2}}$ & $\mathit{0.427}$ & $\mathit{0.501}$ & $\mathit{0.637}$ & $\mathit{1.35}$ \\
		& & & & & \\
		$\mathit{\;\;\;\;\;\;\;\;\;\;\:231\:\left(\bar{1}32\right)}$ & $\mathit{\frac{1}{2}3\frac{3}{2}}$ & $\mathit{0.497}$ & $\mathit{0.749}$ & $\mathit{0.593}$ & $\mathit{2.5}$ \\
		$332$ & $\frac{1}{2}3\frac{5}{2}$ & $0.538$ & $0.749$ & $0.569$ & $2.2$ \\
		$\bm{331}$ & $\bm{132}$ & $\bm{0.549}$ & $\bm{3.929}$ & $\bm{0.563}$ & $\bm{348}$\\
		$330$ & $\frac{3}{2}3\frac{3}{2}$ & $0.577$ & $0.749$ & $0.548$ & $1.95$ \\
		$\mathit{\;\;\;\;\;\;\;\;\;\;\:431\:\left(\bar{1}34\right)}$ & $\mathit{\frac{3}{2}3\frac{5}{2}}$ & $\mathit{0.613}$ & $\mathit{0.749}$ & $\mathit{0.529}$ & $\mathit{1.75}$ \\
		\hline
	\end{tabular}
\caption{\textbf{ Results from frozen phonon model.} Calculations based on an atomic displacement of $\left|\delta\right|=0.01$ along $\bm{b}_{\mbox{\scriptsize s}}$ for the clockwise domain with $\bm{b}_{\mbox{\scriptsize s}}\equiv \bm{b}_{\mbox{\scriptsize p}}$, see section IV for transformation matrices. \textit{Cmcm} and \textit{bcc} indices are given in columns one and two, the \textit{bcc} Bragg positions are bold black, clockwise \textit{Cmcm} positions in normal font, and anticlockwise \textit{Cmcm} positions in italics with correct indices in parentheses. The structure factor, which for diffuse positions depends only on $k_{\mbox{\scriptsize s}} \equiv k_{\mbox{\scriptsize p}}$, is given in column four and column five gives the scattering factor for an averaged atom $\mbox{U}_{\scriptsize 0.75}\mbox{Mo}_{\scriptsize 0.25}$, normalized to $1$ at $\sin\theta/\lambda=0$. The top panel contains calculated results for the reflections in Figure $2$ - main text. The lower panel highlights a selection of diffuse positions around their \textit{bcc} Bragg positions with a focus on higher $\bm{Q}$. Final intensities are normalized to the strongest Bragg reflection $\mbox{I}\left(110\right)_{\mbox{\scriptsize p}}=1000$ with no Lorentz correction applied. Final diffuse intensities are divided by $6$ to account for the reduction in scattering volume resulting from $6$ possible domains.}
\label{Tab1}
\end{table}
\clearpage

\section{Parameter extraction from diffuse peaks}
\begin{figure}[h]
	\centering
	\includegraphics*[scale=1.05]{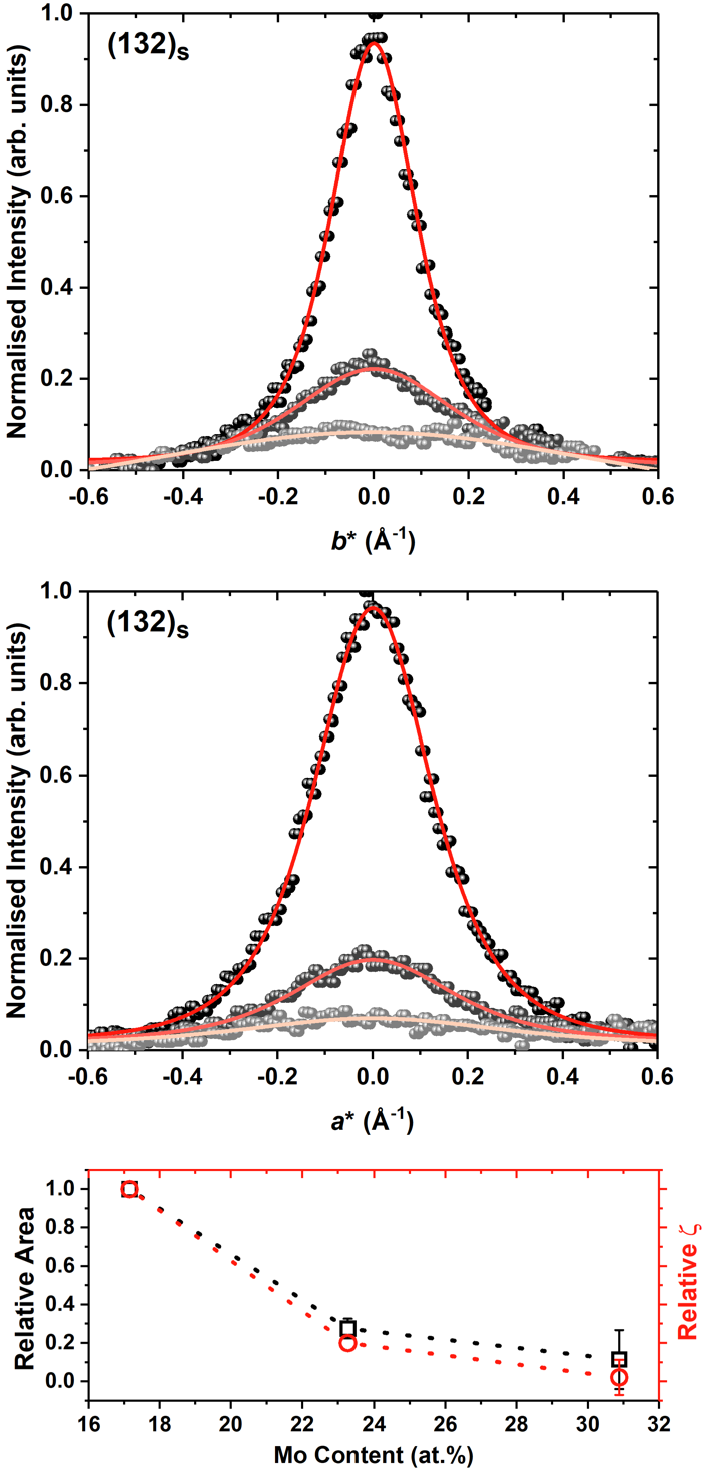}
	\caption{\textbf{ Line profiles across the semi-minor (upper) and semi-major (middle) elliptical axes for the $\bm{(132)_{\mbox{\scriptsize s}}}$ reflection.} The elliptical axes correspond to $\bm{b^*}$ and $\bm{a^*}$ directions, respectively. All three Mo concentrations are shown: $17.2$ at.\% (black), $23.3$ at.\% (dark grey) and $30.9$ at.\% (light grey). Curves correspond to squared Lorentzian best fits. Peak intensity is normalized to the $17.2$ at.\%  maximum and all curves are shifted to set the peak centre equal to zero. Lower panel: Normalized peak area (black) and normalized correlated volume $\zeta=\xi_{b^*}^2 \times \xi_{a^*}$ (red) plotted against Mo concentration. Dotted lines are guides to the eye.  } 
	\label{FigS2}
\end{figure}

Diffuse peak width and integrated intensity were determined via line profiles extracted from reconstructed planes such as Figure $2$ - main text. Pixel width was determined from the known Bragg reflection spacing. As explained in the main text, the peak width may be related to the correlation length $\left(\xi\right)$ of the superstructure as the  $\mbox{Full Width Half Maximum}\left(\mbox{FWHM}\right)=2\pi/\xi$\cite{FWHM} whereas the integrated intensity is proportional to the magnitude of the atomic displacement, $\left|\delta\right|$, along $\bm{b}_{\mbox{\scriptsize s}}$. As there are multiple convoluted sources of broadening present in the diffuse peak this analysis only provides a lower limit on the correlation length. However, given the magnitude of the observed widths, the finite correlation length is likely the dominant effect and therefore the value obtained is expected to be a good approximation for the true value. Figure \textcolor{blue}{S}\ref{FigS2} shows extracted line profiles across the long axis of the $\left(132\right)_{\mbox{\scriptsize s}}$ reflection, highlighted with a blue ellipse in Figure $2$ - main text. As stated in the main paper, the exact character of the correlations is undetermined, however a squared Lorentzian function proved a marginally better fit therefore this form was used for all curves.

An analysis of the line profiles gives correlation lengths (for $17$ at.\% Mo) of $\xi=22\:\mbox{\AA}$ along $\bm{a}_{\mbox{\scriptsize s}}$, and $\xi=30\:\mbox{\AA}$ along $\bm{b}_{\mbox{\scriptsize s}}$ and $\bm{c}_{\mbox{\scriptsize s}}$. As shown in the lower panel, all correlation lengths decrease with increasing Mo content. The elliptical eccentricity is also observed to decrease from a maximum of $0.69$ with increasing alloy content. The integrated intensity, proportional to $\left|\delta\right|$, drops rapidly with increasing Mo concentration from a maximum $\left|\delta\right|$ of between $0.01$ and $0.03$, equivalent to between $0.03$ and $0.1\:\mbox{\AA}$. These values are in line with those found for a $20$ at.\% doped alloy\cite{Yakel}.

The observation that both the FWHM and integrated intensity decay dramatically with increasing alloy content highlights the role played by Mo in stabilizing the global \textit{bcc} structure. We suggest a possible mechanism, specific to the UMo system, could be broadening of the $f\mbox{-band}$, through both added $d\mbox{-character}$ and alloy broadening\cite{Band_strucutre}, reducing the potential energy gain from a Peierls-like distortion\cite{Soderlind}, however the difference in orbital shape between $d$ and $f$ elements may also be relevant\cite{Mettout}. Derived parameters are shown in the lower panel of Figure \textcolor{blue}{S}\ref{FigS2} from which it can be seen that the integrated area and correlated volume decrease in tandem. This disagrees with the model proposed by Starikov \textit{et al.}\cite{Starikov}, who proposed a fixed displacement magnitude with decaying correlations for increased Mo content. Ultimately, they suggest the system enters a quasi-\textit{bcc} state, where each \textit{bcc} unit cell contains an atomic displacement, but there exist no correlations between unit cells. They also predict that a distorted superstructure should only exist over a narrow range of alloy compositions, $\sim16-18\:\mbox{at.\%}$ at $300\:\mbox{K}$. However, in contradiction with their predictions, we observe a \textit{Cmcm} superstructure from $17-31\:\mbox{at.\%}$. It should be noted that their work was premised on the previously proposed structure\cite{Yakel}, however it is unclear if this would explain the discrepancies given the two structures are remarkably similar.
\clearpage
\clearpage

\section{Transformation matrices}
The relevant matrices for transforming Miller indices from a \textit{bcc} representation to any one of the 6 possible positive \textit{Cmcm} domains and back again are given below, such that

\begin{equation*}
\mbox{\textbf{\Large T}}\colvec{3}{h}{k}{l}_{\mbox{\large p}} = \colvec{3}{h}{k}{l}_{\mbox{\large s}} \mbox{\Large and}  \: \: \mbox{\textbf{\Large T}}^{\bm{-1}}\colvec{3}{h}{k}{l}_{\mbox{\large s}} = \colvec{3}{h}{k}{l}_{\mbox{\large p}}\Large .
\end{equation*}

\noindent Matrices for time-reversal pairs are obtained by multiplication by $-1$. 

\vspace*{5mm}
{\centering
	\renewcommand{\arraystretch}{1.25}
	\begin{tabular}{ c c c c c}
	\multicolumn{2}{c}{$\bm{\mbox{\textbf{T}}(bcc-Cmcm)}$} & \hspace*{20mm} & \multicolumn{2}{c}{$\bm{\mbox{\textbf{T}}^{-1}(Cmcm-bcc)}$}\\
	\multicolumn{2}{c}{$\bm{b_{\mbox{\scriptsize s}} \equiv a_{\mbox{\scriptsize p}}}$}	& & \multicolumn{2}{c}{$\bm{b_{\mbox{\scriptsize s}} \equiv a_{\mbox{\scriptsize p}}}$}\\[-10pt]
	\textbf{clockwise}&\textbf{anti-clockwise}& &$~\:\:\mbox{\textbf{clockwise}}$&$~\:\:\:\mbox{\textbf{anti-clockwise}}$\\
	$\boldmath{\setlength{\arraycolsep}{5.6pt}\begin{bmatrix}[0.9] 0 & -1 & 1 \\ 1 & 0 & 0 \\ 0 & 1 & 1 \end{bmatrix}}$ & $\boldmath{\begin{bmatrix}[0.9] 0 & -1 & -1 \\ 1 & 0 & 0 \\ 0 & -1 & 1 \end{bmatrix}}$ & & $\boldmath{\mbox{\LARGE$\frac{1}{2}$}\setlength{\arraycolsep}{5.6pt}\begin{bmatrix}[0.9] 0 & 2 & 0 \\ -1 & 0 & 1 \\ 1 & 0 & 1 \end{bmatrix}}$ & $\boldmath{\mbox{\LARGE$\frac{1}{2}$} \begin{bmatrix}[0.9] 0 & 2 & 0 \\ -1 & 0 & -1 \\ -1 & 0 & 1 \end{bmatrix}}$ \\[25pt]
	\multicolumn{2}{c}{$\bm{b_{\mbox{\scriptsize s}} \equiv b_{\mbox{\scriptsize p}}}$}	& & \multicolumn{2}{c}{$\bm{b_{\mbox{\scriptsize s}} \equiv b_{\mbox{\scriptsize p}}}$}\\[-10pt]
	\textbf{clockwise}&\textbf{anti-clockwise}& &$~\:\:\mbox{\textbf{clockwise}}$&$~\:\:\:\mbox{\textbf{anti-clockwise}}$\\
	$\boldmath{\setlength{\arraycolsep}{5.6pt}\begin{bmatrix}[0.9] 1 & 0 & 1 \\ 0 & 1 & 0 \\ -1 & 0 & 1 \end{bmatrix}}$ & $\boldmath{\setlength{\arraycolsep}{5.6pt}\begin{bmatrix}[0.9] 1 & 0 & -1 \\ 0 & 1 & 0 \\ 1 & 0 & 1 \end{bmatrix}}$ & & $\setlength{\arraycolsep}{5.6pt}\boldmath{\mbox{\LARGE$\frac{1}{2}$}\begin{bmatrix}[0.9] 1 & 0 & -1 \\ 0 & 2 & 0 \\ 1 & 0 & 1 \end{bmatrix}}$ & $\boldmath{\mbox{\LARGE$\frac{1}{2}$}\setlength{\arraycolsep}{5.6pt}\begin{bmatrix}[0.9] 1 & 0 & 1 \\ 0 & 2 & 0 \\ -1 & 0 & 1 \end{bmatrix}}$ \\[25pt]
	\multicolumn{2}{c}{$\bm{b_{\mbox{\scriptsize s}} \equiv c_{\mbox{\scriptsize p}}}$}	& & \multicolumn{2}{c}{$\bm{b_{\mbox{\scriptsize s}} \equiv c_{\mbox{\scriptsize p}}}$}\\[-10pt]
	\textbf{clockwise}&\textbf{anti-clockwise}& &$~\:\:\mbox{\textbf{clockwise}}$&$~\:\:\:\mbox{\textbf{anti-clockwise}}$\\
	$\boldmath{\begin{bmatrix}[0.9] 1 & -1 & 0 \\ 0 & 0 & 1 \\ -1 & -1 & 0 \end{bmatrix}}$ & $\boldmath{\setlength{\arraycolsep}{5.6pt}\begin{bmatrix}[0.9] 1 & 1 & 0 \\ 0 & 0 & 1 \\ 1 & -1 & 0 \end{bmatrix}}$ & & $\boldmath{\mbox{\LARGE$\frac{1}{2}$}\begin{bmatrix}[0.9] 1 & 0 & -1 \\ -1 & 0 & -1 \\ 0 & 2 & 0 \end{bmatrix}}$ & $\boldmath{\mbox{\LARGE$\frac{1}{2}$}\setlength{\arraycolsep}{5.6pt}\begin{bmatrix}[0.9] 1 & 0 & 1 \\ 1 & 0 & -1 \\ 0 & 2 & 0 \end{bmatrix}}$ \\
	\end{tabular}
\par}

\clearpage

\section{Theoretical Methods}
The full theoretical details for the lattice dynamics of the \textit{bcc} U-Mo system have been published recently\cite{Castellano}. All calculations used density-functional theory with the ABINIT package\cite{ABINIT}. Concerning the \textit{ab initio} molecular dynamics (AIMD) simulations, we used a $4\times4\times4$ supercell of $128$ atoms, and the random alloy was modelled using special quasi-random structure methods\cite{SQRS} generated following the method of Van de Walle \textit{et al.}\cite{SQS2} for a 25 at.\% Mo system. The AIMD trajectory was equilibrated during $5-10\:\mbox{ps}$. 

The finite temperature phonon  dispersions and density of states (DOS) were extracted from the AIMD runs using the Temperature Dependent Effective Potential (TDEP) method\cite{TDEP,HT-HP,A-TDEP}. In this method, a set of forces and positions extracted from the AIMD is fitted on a model harmonic Hamiltonian of the form

\begin{equation}
H = U_{\scriptsize 0} + \sum_{\mbox{\scriptsize i}}^{}\frac{p^{2}_{\mbox{\scriptsize i}}}{2m_{\mbox{\scriptsize i}}} + \sum_{\mbox{\scriptsize i,j}}^{} \Theta_{\mbox{\scriptsize i,j}}u_{\mbox{\scriptsize i}}u_{\mbox{\scriptsize j}}
\end{equation}

\noindent with $m_{\mbox{\scriptsize i}}$, $p_{\mbox{\scriptsize i}}$ and $u_{\mbox{\scriptsize i}}$, the mass, momentum and displacement of atom i, respectively. $U_{\scriptsize 0}$ is the ground-state energy and $\Theta_{\mbox{\scriptsize i,j}}$ the second-order effective Interatomic Force Constants (IFC). To model the random alloy while keeping a small number fitting parameters, the symmetry of the underlying structure is imposed on the IFC. 

Within this constraint, two levels of approximation were considered. In the Virtual Crystal Approximation (VCA-TDEP), an averaged, ideal crystal is used during the fit. By preserving the symmetries of the underlying structure, the VCA-TDEP method gives the vibrational properties with standard harmonic formulas. We also used the Symmetry-Imposed Force Constant (SIFC-TDEP) extension of TDEP, where each atom has its real mass, and the IFC are type dependent. For more general details see ref.~\citenum{Shulumba}, and for a more detailed explanation of the approximations see ref.~\citenum{Castellano}.

The SIFC-TDEP method breaks the underlying structural symmetry, so the supercell must be considered to be the unit cell during the phonon extraction. Although this does not affect the phonon-DOS extraction, the resulting dispersion cannot be compared directly with experimental results. To obtain the theoretical dispersion shown in the main text, we used the band unfolding method of Ikeda \textit{et al.}\cite{Ikeda} In this unfolding procedure, the spectral function at a given $k$-point is computed by projecting the supercell phonon dispersion onto the underlying structure. One should note that with this method, the calculated linewidth broadening is only due to the variation in mass and IFC's, neither phonon-phonon interactions (i.e. anharmonic contributions) nor correlated disorder are included in the estimated linewidth. Anharmonic contributions are present during AIMD simulations, and are extracted in an “effective way” using the TDEP method. Via a separate method we have assessed the linewidth contribution from phonon-phonon interactions to be much less than $1\:\mbox{meV}$, as expected at room temperature\cite{BZ_lifetimes}. 
\clearpage

\section{Theoretical Phonon Density of States}
\begin{figure}[h]
	\centering
	\includegraphics*[scale=0.0645]{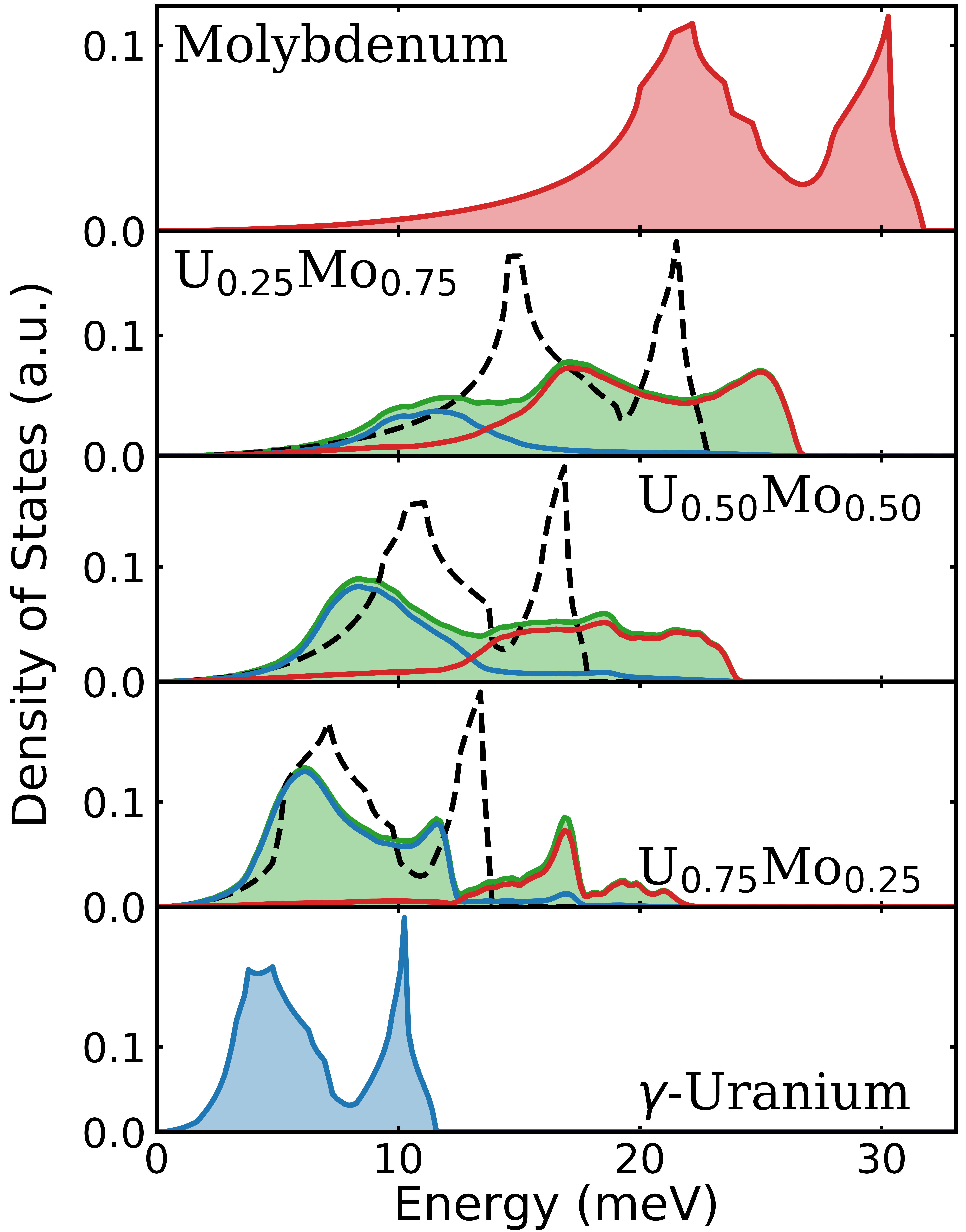}
	\caption{\textbf{ Computed phonon density of states for various U-Mo alloys at $\bm{900}\:\mbox{K}$.} The dashed and full lines are the DOS computed with the VCA and SIFC methods, respectively. Molybdenum and uranium contributions are shown in red and blue respectively while the total DOS is shown in green.} 
	\label{FigS3}
\end{figure}

Figure \textcolor{blue}{S}\ref{FigS3} shows theoretical phonon-DOS for the \textit{bcc} U-Mo system, at $900\:\mbox{K}$, with various Mo concentrations. Within the VCA, the averaging of both atomic masses and IFC's results in an ideal crystal and consequently, the phonon-DOS display sharp peaks. The inclusion of mass disorder through SIFC separates the phonon DOS into two broadened frequency domains: a lower frequency domain dominated by uranium excitations and a higher frequency domain dominated by molybdenum. The broadening of the phonon-DOS in the SIFC method, as compared to the VCA model, is a direct consequence of considering the mass and IFC fluctuations present in the alloy. Some of the authors have recently shown that this broadening is responsible for the stabilization of the $\gamma$ (\textit{bcc}) U-Mo phase at $900\:\mbox{K}$\cite{Castellano}. The higher energy modes of molybdenum, above $15\:\mbox{meV}$, are difficult to observe experimentally for a variety of reasons. Firstly, molybdenum is the minor alloy component and, secondly, molybdenum has both a lower scattering factor compared to uranium, and much smaller x-ray inelastic scattering cross section due to the relevant weighting by inverse energy.  

\clearpage

\section{Transverse acoustic phonon softening in the vicinity of the H position}
\begin{figure}[h]
	\centering
	\includegraphics*[width=0.9\linewidth]{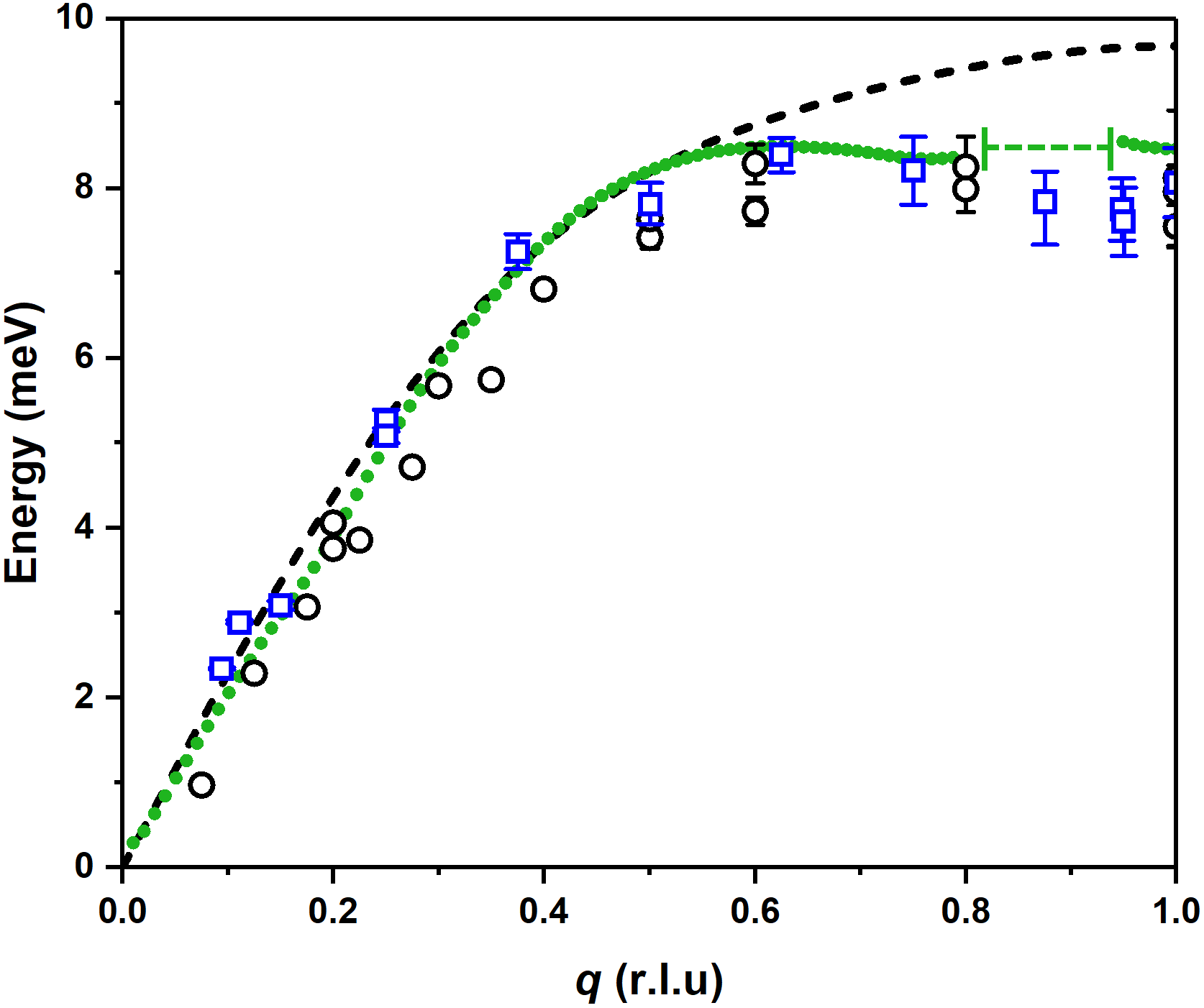}
	\caption{\textbf{ TA softening near the H position.} Experimental points are shown as open blue squares with errors estimated using a threshold $10\%$ increase in the total least squared error. The error in the phonon energies and linewidths were assumed to be independent. Open black circles are data reproduced from Brubaker \textit{et al.}\cite{LLNL} The predicted TA dispersion, as given by the VCA method, is shown by a dashed black line, scaled by $1.24$ to provide good agreement with experimental data in the long-wavelength limit. The dotted green line is the estimate from the SIFC method (See Sec. V, VI and IX), with the “anomalous region” shown as a dashed green section. } 
	\label{FigS5}
\end{figure}

We observe a moderate softening for the $\mbox{TA}\left[001\right]_{\mbox{\scriptsize p}}$ mode near the zone boundary, as shown clearly in Figure \textcolor{blue}{S}\ref{FigS5}. This softening was also observed by Brubaker \textit{et al.} and tentatively attributed to either the metastability of the \textit{bcc} phase, strong electron-phonon coupling, or a combination of both\cite{LLNL}. To evaluate the degree and origin of the softening we scaled the theoretical dispersion produced by both the virtual crystal approximation (VCA) (dashed black line) and the symmetry-imposed force constant (SIFC) approach (dotted green line; see Sec. V, VI \& IX), such that the theoretical and experimental results agree in the long-wavelength limit. This scaling was designed to account for the systematic underprediction observed between theory and experiment across the entire dispersion in Figure $5$ - main text, likely due to the complexities in modelling $5f$ systems. 

The maximum softening may then be extracted as roughly $2\:\mbox{meV}$ by comparing the experimental results with the VCA prediction. The electronic effects of alloying are captured at the \textit{ab initio} stage and are therefore included in both the VCA and SIFC results. However the SIFC approach, which appears to accurately predict the softening of this mode from $\bm{q}\sim0.6$ to the zone boundary, also accounts for the spatial fluctuations in both atomic mass and IFC’s that are induced by alloying. This good agreement suggests that the origin of the softening is a direct alloying effect and, in contrast to the speculation of Brubaker \textit{et al.}\cite{LLNL}, is not primarily driven by either crystal or electronic structure. It should be noted that where the LA and TA modes are close to degenerate, around $\bm{q}=0.85$ (see Fig. 5 - main text), there is uncertainty in mode assignment when extracting energies from the spectral function as information on the underlying crystallographic symmetry is lost when a disordered system is mimicked by a supercell model\cite{Ikeda}.
\clearpage
\vspace*{20mm}
\clearpage

\section{Assessing the effect of length dependent interatomic force constants on the spectral function at \texorpdfstring{$\bm{q =\mbox{\nicefrac{2}{3}} \left[111\right]_{\mbox{\scriptsize p}}}$}{q}}

\begin{figure}[h]
	\centering
	\includegraphics*[width=\linewidth]{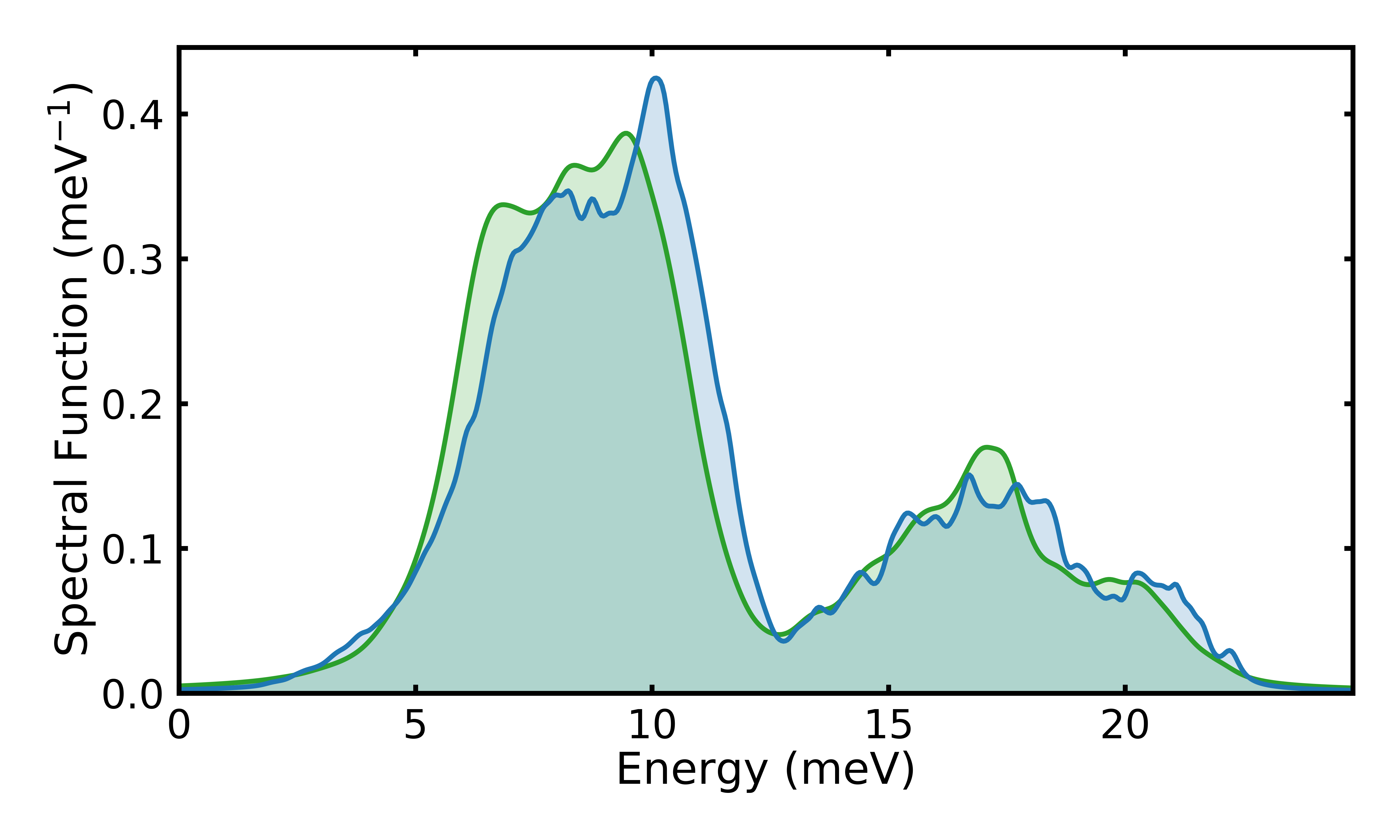}
	\caption{\textbf{ Theoretical spectral function at $\bm{q =\mbox{\nicefrac{2}{3}} \left[111\right]_{\mbox{\scriptsize p}}}$ at $\bm{300}\:\mbox{K}$.} Results inclusive and exclusive of length dependent interatomic force constants shown as blue and green curves, respectively.} 
	\label{FigS6}
\end{figure}

Figure \textcolor{blue}{S}\ref{FigS6} shows the resulting spectral functions inclusive and exclusive of length dependent interatomic force constants using the method described in the main text. In both cases the lower energy intensity is formed from two peaks that correspond to the LA (lower energy) and TA (higher energy) modes whereas the higher energy intensity is predominantly due to Mo-Mo modes. As stated in the main text both the LA and TA modes are shifted upwards in energy by roughly $1\;\mbox{meV}$. There may also be an upwards shift in the Mo-Mo modes, however the width of the spectral response for these modes makes it difficult to distinguish.

\clearpage